\documentclass[usenatbib]{mn2e}
\usepackage{amssymb, amsmath}
\usepackage{epsfig}
\usepackage{graphicx}
\title{Substructure in the Most Massive GEEC Groups: Field-like Populations in Dynamically Active Groups}
\author[Hou et al.]{Annie Hou$^{1}$, Laura C. Parker$^{1}$, David J. Wilman$^{2}$, Sean L. McGee$^{3,4}$, William E. Harris$^{1}$, \newauthor Jennifer L. Connelly$^{2}$, Michael L. Balogh$^{4}$, John S. Mulchaey$^{5}$, Richard G. Bower$^{3}$\\
$^{1}$Department of Physics $\&$ Astronomy, McMaster University, Hamilton ON L8S 4M1, Canada\\
$^{2}$Max-Planck-Institut f$\ddot{u}$r Extraterrestrische Physik, Giessenbachstra$\ss$e,  D-85748 Garching, Germany\\
$^{3}$Department Of Physics, University of Durham, Durham, UK DH1 3LE\\
$^{4}$Department of Physics and Astronomy, Univeristy of Waterloo, Waterloo, Ontario, N2L 3G1, Canada\\
$^{5}$Observatories of the Carnegie Institution, 813 Santa Barbara Street, Pasadena, California, USA}

\begin{document}
 \maketitle
\begin{abstract}
The presence of substructure in galaxy groups and clusters is believed to be a sign of recent galaxy accretion and can be used not only to probe the assembly history of these structures, but also the evolution of their member galaxies.  Using the Dressler-Shectman (DS) Test, we study substructure in a sample of intermediate redshift ($z \sim 0.4$) galaxy groups from the Group Environment and Evolution Collaboration (GEEC) group catalog.  We find that 4 of the 15 rich GEEC groups, with an average velocity dispersion of $\sim$525 km s$^{-1}$, are identified as having significant substructure.  The identified regions of localized substructure lie on the group outskirts and in some cases appear to be infalling.  In a comparison of galaxy properties for the members of groups with and without substructure, we find that the groups with substructure have a significantly higher fraction of blue and star-forming galaxies and a parent colour distribution that resembles that of the field population rather than the overall group population.  In addition, we observe correlations between the detection of substructure and other dynamical measures, such as velocity distributions and velocity dispersion profiles.  Based on this analysis, we conclude that some galaxy groups contain significant substructure and that these groups have properties and galaxy populations that differ from groups with no detected substructure.  These results indicate that the substructure galaxies, which lie preferentially on the group outskirts and could be infalling, do not exhibit signs of environmental effects, since little or no star-formation quenching is observed in these systems.
\end{abstract}

\section{Introduction}
The current theory of structure formation in the Universe is based on the standard $\Lambda$ cold dark matter ($\Lambda$CDM) cosmological model, in which objects grow hierarchically from the initial matter density perturbations through mergers and accretion \citep[e.g.,][]{ps74, lc93, springel05}.  In order to test the theory of hierarchical structure formation, one must investigate the assembly history of the large structures in the Universe, namely galaxy groups and clusters. 

A natural consequence of a hierarchical Universe is the existence of substructure within larger systems.  Traditionally, substructure has been defined as a kinematically distinct sub-halo within a larger parent halo.  A broader, more observationally motivated definition, and one that we will assume here, also includes separate haloes that are either merging to form a larger halo; gravitationally bound and infalling onto a pre-existing halo; or in the nearby large-scale-structure, but not necessarily bound or infalling.  In general, accreting structure does not have the same kinematic properties as the host, but whether or not substructure can be observed depends on how long it remains intact after infall.  Early $N$-body simulations suggested that the assimilation of substructure into the host was rapid, providing no detectable long-lived features \citep{kw93, summers95}.   However, later work has shown that the lack of observable substructure in these simulations was due to poor resolution \citep{moore96}.  Indeed, high resolution $N$-body simulations \citep[e.g.,][]{diemand08,springel08}, demonstrate that several hundred thousand sub-haloes can exist in a Milky Way sized dark matter halo at a redshift of zero.  Using semi-analytic models to study the substructure within individual galaxy to cluster-sized haloes, \citet{taylor04} showed that accreting systems could survive within the host halo for many orbits, depending on the orbital parameters of the substructure upon infall.  

These theoretical results suggest that substructure should be a detectable quantity and numerical dark matter simulations of galaxy groups and clusters in a $\Lambda$CDM Universe predict that approximately 30 per cent of all systems should contain substructure \citep{knebe2000}.  Studies of individual clusters \citep[e.g.,][]{beers82, ds88, wb90, bird94, cd96, burgett04, bohringer10} indicate that a large fraction of systems show evidence of significant sub-clustering.  The predicted theoretical value of 30 per cent is in agreement with some substructure studies of groups and clusters \citep[e.g.,][]{zm98a, solanes99}, but several other results have demonstrated a much higher fraction of substructure; \citet{bird94b} observed that 44 per cent\footnote{11/25 galaxies in their sample have significant ($>$ 95 per cent) substructure based on results of the Dressler-Shectman statistics.} of their sample contained substructure, \citet{ds88} observed 53 per cent and \citet{ramella07} find an extremely high substructure fraction of 73 per cent.  Although the precise fraction of groups and clusters with substructure may still be a source of debate, the observed presence of any substructure strongly suggests that these systems grow in a hierarchical manner through the accretion of galaxies and smaller groups of galaxies.

Though galaxy clusters are amongst the largest structures in the local Universe, they do not represent the most common host environment for galaxies.  Galaxy groups, which contain a few to tens of member galaxies, are the host environment of more than half of the present-day galaxy population \citep{gh83, eke05}.  Despite the importance of groups in the build up of large galaxy clusters, there have been few studies on the assembly history of groups themselves.  

One method of probing group assembly histories is to look at the amount of substructure within these systems.  The presence of such structure would indicate that the group has recently accreted galaxies  \citep{lc93}.  Studies have been carried out for galaxy groups in the local Universe by \citet{zm98a}, who observed significant ($>$ 99 per cent confidence level) substructure in two of their six local groups.  An interesting result of their analysis showed that the substructure was located on the outskirts of the systems, that is $\sim$0.3-0.4 $h^{-1}$ Mpc, where ${h = H_{0}} /(100$ km s$^{-1}$ Mpc$^{-1}$), from the core of the group.  Based on these results, \citet{zm98a} concluded that like rich galaxy clusters, some galaxy groups assembled in a hierarchical manner through the accretion of smaller structures from the field.  A more recent study of local galaxy groups by \citet{firth06} found similar results, with roughly half of their sample showing significant substructure.  Although, the findings of \citet{zm98a} and \citet{firth06} provided an important first step in unveiling the assembly history of groups, their analysis was based on a small sample of very nearby systems.  In order to gain a better understanding of the role of groups in the growth of structure, we must search for substructure in a larger sample of galaxy groups with highly complete spectroscopy that cover a wide range in redshift.  Such studies have only become possible with recent deep spectroscopic surveys that have produced large group catalogues, such as the Sloan Digital Sky Survey (SDSS) \citep{berlind06,yang07}, the Group Environment and Evolution Collaboration (GEEC) \citep{wilman05, carlberg01}, and the higher redshift extension of GEEC (GEEC2) \citep{balogh11} optical group catalogs.   

In addition to the role of the group environment in the growth of large scale structure in the Universe, studies of substructure within groups allows us to probe galaxy evolution.  Since groups have fewer members than galaxy clusters, any substructure present will have a stronger effect on the observed group properties (i.e.\ colours, blue or active fractions).  If substructure traces accreting galaxies, one might expect a correlation between observed galaxy properties and substructure.  Possible correlations could exist between substructure and the colours of galaxies in groups, or with substructure and star formation rates.  One of the main goals of this paper is to search for such correlations. 

In this paper we search for substructure in a sample of intermediate redshift galaxy groups from the GEEC Group Catalog.  In Section 2, we discuss the sensitivity and reliability of the Dressler-Shectman (DS) Test for group-sized systems using Monte Carlo simulations.  In Section 3, we apply the DS Test to the GEEC groups and discuss the results of our analysis.  In Section 4, we discuss the relationship between the presence of substructure and other indicators of dynamical state, such as the shape of the group velocity distribution.  In Section 5, we look for correlations between substructure and the properties of members galaxies, such as colour and star formation rate, and discuss the possible implications of our findings.  In Section 6 we present our conclusions.  Additionally, we include an Appendix, which provides detailed results of our Monte Carlo Simulations.

Throughout this paper we assume a $\Lambda$CDM cosmology with $\Omega_{M} = 0.3$, $\Omega_{\Lambda} = 0.7$, and $H_{0}$ = 75 km s$^{-1}$ Mpc$^{-1}$. 

\section{Detecting Substructure in Groups}
Numerous tests for substructure have been developed and carried out for cluster-sized systems \citep[see][for a thorough review]{pinkney96}.   In a comparison of five 3-dimensional (3-D) tests, which use both velocity and spatial information, \citet{pinkney96} determined that the Dressler-Shectman (DS) Test \citep{ds88} was the most sensitive test for substructure, for systems with as few as 30 members.  In the following section we look at the DS Test, and determine its reliability and robustness for smaller group-sized systems. 

\subsection{The Dressler-Shectman (DS) Test}
Substructure manifests itself as detectable deviations in the spatial and/or velocity structure of a system.  The aim of the DS Test is to compute \emph{local} mean velocity and velocity dispersion values, for each individual galaxy and its nearest neighbours, and then compare these to the global group values.  Following the notation of \citet{ds88}, we define $(\bar{\nu}, \sigma)$ as the mean velocity and velocity dispersion of the entire group, which is assumed to have $n_{\rm{members}}$ galaxies.  Then for each galaxy $i$ in the group, we select it plus a number of its nearest neighbours, $N_{\rm{nn}}$, and compute their mean velocity $\bar{\nu}^i_{\rm{local}}$ and velocity dispersion $\sigma^i_{\rm{local}}$.  From these we compute the individual galaxy deviations, $\delta_{i}$, given as
\begin{equation}
\centering
\delta_{i}^{2} = \left(\frac{N_{\text{nn}} + 1}{\sigma^{2}}\right)\left[\left(\overline{\nu}^i_{\text{local}} - \overline{\nu}\right)^{2} + \left(\sigma^i_{\text{local}} - \sigma\right)^{2}\right],
\label{delis}
\end{equation}
\\
\noindent where $1 \leq i \leq n_{\rm{members}}$ (i.e.\  $\delta_{i}$ is computed for each galaxy in the system).  \citet{ds88} originally developed the test for cluster-sized systems and the number of nearest neighbours used to compute the $\delta_{i}$ values was relatively high (i.e.\ $N_{\text{nn}}$ = 11).  Since substructure in galaxy groups is likely to have fewer than 11 constituent galaxies, we take $N_{\text{nn}}$ to be $\sqrt{n_{\rm{members}}}$ following the methodology of authors who have previously applied the DS Test to group-size systems \citep{silverman86, pinkney96, zm98b}.  This ensures that large kinematic deviations of a few neighbouring galaxies do not get diluted by adding too many unassociated galaxies, and thereby lowering the computed $\overline{\nu}^i_{\text{local}}$ and $\sigma^i_{\text{local}}$ values in Equation \ref{delis}.

The statistic used in the DS Test is the $\Delta$-value, given by
\begin{equation}
\centering
\Delta = \sum_{i}\delta_{i}.
\label{Delta}
\end{equation}
\\
A system is then considered to have substructure if $\Delta/n_{\rm{members}} > 1.0$ \citep{ds88}.  This method of using a threshold value to find substructure is referred to as the critical values method.

An alternative method of identifying substructure with the DS Test is to use probabilities ($P$-values) rather than critical values.  The $P$-values for the DS Test are computed by comparing the observed $\Delta$-value to `shuffled' $\Delta$-values, which are computed by randomly shuffling the observed velocities and re-assigning these values to the member positions, a procedure called `MC shuffling'.  The $P$-values are given by
\begin{equation}
\centering
P = \sum \left(\Delta_{\text{shuffled}} > \Delta_{\text{observed}}\right)/n_{\text{shuffle}}.
\label{DSprob}
\end{equation}
\\
where $\Delta_{\text{shuffled}}$ and $\Delta_{\text{observed}}$ are both computed using Equation \ref{Delta} and $n_{\text{shuffle}}$ is the number of `MC shuffles', and therefore the number of $\Delta_{\text{shuffled}}$-values, used to compute the probability .  One can see from Equation \ref{DSprob} that systems with significant substructure will have low $P$-values, since it is unlikely to obtain the observed $\Delta$-value randomly (Equation \ref{Delta}). 

\subsection{Monte Carlo Simulations}
Although \citet{pinkney96} have carried out an extensive investigation of the DS Test, their analysis was performed on systems with a minimum of 30 members, and the majority of our intermediate redshift groups have fewer member galaxies (see Section 3.1).  Therefore, we perform our own Monte Carlo simulations in order to check the reliability, sensitivity and robustness of the DS Test, using both the critical value and probability ($P$-value) methods, for group-sized systems.  It should be noted that we specifically model our host mock groups after the observed GEEC group sample.

\subsubsection{Generating the Mock Groups}
Mock galaxy groups, both with and without substructure, are generated using Monte Carlo methods to assign member galaxy positions.  These mock groups are then used to compute the false negative and positive rates of the DS Test, and also determine the sensitivity of the test against a variety of input parameters.

\begin{figure}
\includegraphics[width = 8cm, height = 8cm]{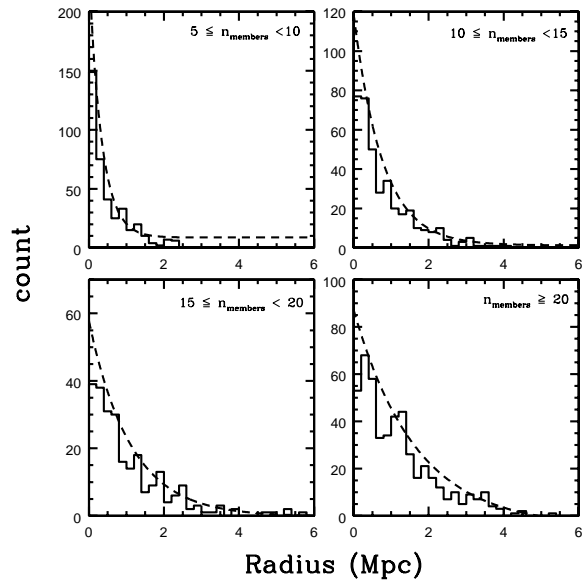} 
\caption{Histograms of the radial distribution of the galaxies in the GEEC Group Catalog stacked in bins of $n_{\rm{members}}$.  The dashed line corresponds to an exponential fit to the distribution, which we use to generate the radial positions of galaxies in our mock groups. \label{radhist}}
\end{figure}

The radial positions for the members of the mock groups are randomly drawn from fits to the projected group-centric radial distributions of the galaxies observed in the GEEC group catalog \citep{wilman05}.  The groups are divided into four bins based on the number of members in each group (Fig. \ref{radhist}), and fits to each bin are used to populate the mock groups.  Since the GEEC groups span such a wide range of masses and group membership, we elect not to use a single fit to the radial distribution of all the group galaxies in our sample.  Instead, we divide our sample into bins of group membership ($5 \leq n_{\rm{members}} < 10$, $10 \leq n_{\rm{members}} < 15$, $15 \leq n_{\rm{members}} < 20$ and $n_{\rm{members}} \geq 20$) and fit each distribution separately.  It should be noted that we bin our sample by group membership, rather than mass or velocity dispersion, as we aim to study the false positive and negative rates of the DS Test as a function of $n_{\rm{members}}$.  However, the results are similar if we bin by $\sigma$ rather than $n_{\rm{members}}$.  The histograms and fits to the radial distributions of the binned groups are shown in Fig. \ref{radhist}.  The general form of the radial distribution of member galaxies is given by

\begin{equation}
\centering
N \propto \exp\left(-\lambda R\right),
\label{raddist}
\end{equation}
\\ 
where  $R$ is the radial position of the given galaxy and $\lambda$ = 2.98, 1.31, 0.902 and 0.606 Mpc$^{-1}$ for the $5 \leq n_{\rm{members}} < 10$, $10 \leq n_{\rm{members}} < 15$, $15 \leq n_{\rm{members}} < 20$ and $n_{\rm{members}} \geq 20$ bins, respectively.  From these results it is clear that groups with fewer members have steeper radial distributions and smaller maximum group centric radii (Fig. \ref{radhist}).

The redshifts of the member galaxies in the mock groups are randomly drawn from Gaussian distributions.  The dispersions of the input Gaussians are taken to be the average velocity dispersion of the aforementioned group membership bins, which are: 300, 350, 400 and 550 km s$^{-1}$.  These dispersion values are chosen for the mocks groups in order to best mimic the systems in our GEEC sample.  We also generate mock groups with the same input host velocity dispersion ($\sigma_{\text{host}}$) for all values of $n_{\rm{members}}$ and find no significant difference in our results (see Appendix for further detail).

As will be shown in Section 2.2.2, in order to determine the false negative rates for the DS Test, we must include substructure within the host group.  The positions and redshifts of the substructure galaxies are drawn from a Gaussian distribution\footnote{Tests with non-Gaussian input substructure have also been carried out and for all reasonable distributions the results do not differ significantly from the results presented here.}.  The free parameters in our mock groups with substructure are: the velocity dispersion of the substructure ($\sigma_{\text{redshift}}$), spatial dispersion of the substructure ($\sigma_{\text{position}}$), location of the substructure, in both position ($\epsilon_{\text{position}}$) and velocity ($\epsilon_{\text{redshift}}$) space, the number of galaxies in the substructure ($n_{\text{sub}}$), the number of galaxies in the host group ($n_{\rm{members}}$), and velocity dispersion of the host group ($\sigma_{\text{host}}$).  We briefly analyze the effects of each of these parameters on the robustness of the DS Test in the following sections, and give a more detailed discussion in the Appendix.

\subsubsection{Testing for False Positives}
Using the mock groups with no input substructure described in Section 2.2.1, we determine the false positive rates, or type I errors, which for the DS Test occur when substructure is identified when none exists.  The only parameter we vary for these systems is the number of member galaxies ($n_{\rm{members}}$).  

We first compute the false positive rates using the critical values method.  For each value of $n_{\rm{members}}$ we compute the $\Delta$-statistic, (Equations \ref{delis} and \ref{Delta}) and then classify a group as having substructure if $\Delta/n_{\rm{members}} > 1.0$.  With this criterion, we find that for all relevant values of $n_{\rm{members}}$ the false positive rates are extremely high.  For example for $n_{\rm{members}} = 20$, the false positive rate is $\sim$81 per cent and even for larger systems, $n_{\rm{members}} = 100$, we find that the rate is equally high.  A similar result was observed by \citet{knebe2000} in a study of substructure in numerically simulated galaxy clusters.  They found that for haloes with no substructure the $\Delta/n_{\rm{members}}$- values were often greater than 1.0, with values peaking closer to 1.4 for larger (i.e.\ $n_{\rm{members}} \sim$80 -100) clusters.  Although it is possible to better identify substructure in richer systems, such as clusters, with a higher value of $\Delta/n_{\rm{members}} \sim$1.4 - 1.6 \citep[e.g.,][]{girardi05,knebe2000}, we find that for group-sized systems this methodology could not simultaneously produce low false positive and false negative rates.

Alternatively, when $P$-values are used to identify substructure, we find that the false positive rates are much lower than with the critical values method and also remarkably stable for a wide range of group members (i.e.\ $5 \leq n_{\rm{members}} \leq 50$).  From our mock groups we find that for significance levels of 0.01, 0.05 and 0.10, the false positive rates are 5, 10 and 15 per cent for all values of $n_{\rm{members}}$ tested.  In other words, the DS Test will falsely identify roughly 5 per cent more substructure than the desired significance level.  Although these rates are higher than the expected values for a given significance level, they are still substantially lower than the rates obtained using the critical values method.  Also, as long as one chooses significance levels of 0.01 or 0.05, the rate of false identifications is acceptably low.  Based on these results, we rule out the critical values method and only perform the following analysis using the probabilities, $P$-values, method.  

\subsubsection{Testing for False Negatives}
Having determined the rate of false identifications obtained with the DS Test, we now test for the rate of false negatives, or type II errors.  For this statistic the false negative rate measures how often the DS Test fails to detect substructure when indeed it exists.  The purpose of conducting these false negative tests is twofold: first, they help determine the reliability of the statistic and second, by varying only one of the free input parameters (listed in Section 2.2.1) at a time, you can determine how sensitive the test is to each parameter.  The latter places quantitative constraints on the maximum size and location of substructure that can be identified.  

Before we begin looking at the input parameters, we first determine how reliable the test is at identifying `obvious' substructure, that is to say galaxies that are tightly correlated and located far from the group centre, both in projection on the sky and along the line-of-sight (see Table \ref{basefneg} for input parameter values).  For these mock groups, we find that substructure is correctly identified in almost all cases, using the $P$-value method and  a significance level of 0.01, producing false negative rates of 0 or 1 per cent. 

\begin{table*}
\centering
\caption{Input parameter values for the `base' mock groups described in Section 2.2.3; i.e.\ Monte Carlo groups with a zero false negative rate. \label{basefneg}}
\vspace{0.5cm}
\begin{tabular}{ccccccc}
\hline\hline
$n_{\rm{members}}$ & $n_{\text{sub}}$ & $\sigma_{\text{host}}$ & $\sigma_{\text{position}}$ & $\epsilon_{\text{position}}$ & $\sigma_{\text{redshift}}$ & $\epsilon_{\text{redshift}}$\\
 & & km s$^{-1}$ & Mpc & Mpc & km s$^{-1}$ & km s$^{-1}$\\
\hline
10 & 4 & 350 & 0.01 & 0.5 & 100 & 1300\\
15 & 5 & 400 & 0.01 & 0.5 & 100 & 1300\\  
20 & 5 & 550 & 0.01 & 0.5 & 100 & 1300\\
50 & 10 & 550 & 0.01 & 0.5 & 100 & 1300\\
\hline
\end{tabular}
\end{table*}

We then investigate the individual free parameters in more detail to determine how each alters the false negative rate.  We briefly discuss the main results here and leave the detailed quantitative analysis, as well as full tables of false negative rates, for the Appendix.

Of the five free parameters, we find that the DS Test is most sensitive to the number of galaxies in the substructure ($n_{\text{sub}}$) and the location of the peak of the substructure's velocity distribution with respect to the peak of the host's ($\epsilon_{\text{redshift}}$).  Previous studies of substructure in galaxy clusters by \citet{ds88} and \citet{pinkney96} showed that the DS Test was unable to find substructure that was `superimposed' with the highest density regions of the hosts, since the galaxies were spatially mixed.  \citet{pinkney96} even stated that in these cases, any type of 3-D test would not be able to accurately detect substructure.  An important distinction, not made by these authors, is that two forms of `superposition' can occur.  Substructure can be superimposed with the host group either in projected angular position or in redshift space.  Our analysis shows that the DS Test is significantly more sensitive to $\epsilon_{\text{redshift}}$ than to $\epsilon_{\text{position}}$.  The test can usually identify substructure with superpositions as projected on the sky, but has a very difficult time with those along the line-of-sight.  From Equation \ref{delis}, it is clear that only collections of neighbouring galaxies with a different local mean velocity and velocity dispersion will produce high $\delta_{i}$ values, no matter their angular position.

In addition we also find that the level of sensitivity of the DS Test to $\epsilon_{\text{redshift}}$, the distance between the substructure's and host's peak velocity distribution, is dependent on the number of members in the host group.  We find that for small groups (i.e.\ $n_{\rm{members}} < 20$), the input substructure needs to be further than $2 \sigma_{\text{host}}$ from the group centre in order to be detected.  On the other hand, the DS Test can identify substructure that is roughly $2 \sigma_{\text{host}}$ away from the peak of the host's Gaussian velocity distribution in groups with more than 20 members.  For even larger systems, that is clusters with $n_{\rm{members}} \geq 50$, the input substructure can be located \emph{within} $1 \sigma_{\text{host}}$ and still be identified.  

Another result from this analysis is that the DS Test is sensitive to the number of galaxies that are part of the substructure ($n_{\text{sub}}$).  Our simulations show that no matter the membership of the host group, the test cannot identify substructure with fewer then four members.  Also, the minimum required number of members within the substructure increases with $n_{\rm{members}}$.  This is due to the fact that we set $N_{nn}$ in Equation \ref{delis} to $\sqrt{n_{\rm{members}}}$.  Therefore, more members in the host group means that the velocity information of more `neighbours' will be used to calculate $\overline{\nu}^i_{\text{local}}$ and $\sigma^i_{\text{local}}$.  Thus, if there are too few galaxies in the substructure, their kinematic deviations can be washed out by other neighbouring host galaxies. 

We find that for rich galaxy groups, with $n_{\rm{members}} \geq 20$, the DS Test can reliably identify true substructure, as has been shown by several other authors \citep[e.g.,][]{ds88,pinkney96,knebe2000}.   For systems with fewer member galaxies, such as poor groups with $n_{\rm{members}} < 20$, the false negative rates are very low ($< 5$ per cent) only for tightly correlated substructure galaxies with large kinematic deviations.  In these poor groups, substructure that is more loosely associated with a velocity peak close to that of its host group is not as easily detected by the test.

Taking into account the results of both the false negative and positive tests, we conclude that for systems with $n_{\rm{members}} \geq 20$, the DS Test can reliably identify real substructure if the $P$-values method is employed using a confidence level of either 0.01 or 0.05.  For groups with $n_{\rm{members}}  < 20$, we find that the percentage of groups with substructure identified by the DS Test should be taken as a \emph{lower limit}.

\section{Substructure in the GEEC Groups}
\subsection{The GEEC Group Catalog}
\indent{}The Group Environment and Evolution Collaboration (GEEC) group catalog is based on a set of $\sim$200 intermediate redshift, $0.1 < z < 0.6$, galaxy groups initially identified in the second Canadian Network for Observational Cosmology (CNOC2) redshift survey \citep{yee00, carlberg01}.  The CNOC2 survey observed $\sim$4 $\times 10^{4}$ galaxies covering four patches, totalling 1.5 deg$^{2}$ in area, in the $UBVR_{C}I_{C}$ bands down to a limiting magnitude of $R_{C} = 23.0$.  Spectra of more than 6000 galaxies were obtained with the MOS spectrograph on the Canada-France-Hawaii Telescope (CFHT), with 48 per cent completeness at $R_{C} = 21.5$ \citep{yee00}.  The GEEC group catalog includes extensive follow-up spectroscopy with the Inamori Magellan Areal Camera and Spectrograph (IMACS) (Connelly, J. et al., submitted) and Low Dispersion Survey Spectrograph (LDSS2) on Magellan \citep{wilman05}, as well the Focal Reducer and low dispersion Spectrograph (FORS2) on the Very Large Telescope (VLT) (Connelly, J. et al., submitted).  We have also obtained multi-wavelength imaging data, which includes: X-ray observations with the X-ray Multi-Mirror Mission-Newton (XMM-Newton) and Chandra X-ray Observatories \citep{finog09}, ultraviolet observations with Galaxy Evolution Explorer (GALEX) \citep{mcgee11}, optical observations with the Advanced Camera for Surveys (ACS) on the Hubble Space Telescope (HST) \citep{wilman09}, infrared observations with the Multi-band Imaging Photometer for Spitzer (MIPS) on the Spitzer Space Telescope \citep{tyler11}, and near-infrared observations with Isaac Newton Group Red Imaging Device (INGRID) on the William Herschel Telescope \citep{balogh09}, Infrared Array Camera (IRAC) on Spitzer \citep{wilman08}, and the Son of ISAAC (SOFI) on the New Technology Telescope (NTT) \citep{balogh09}.  In addition, improved optical imaging was obtained in the $ugrizBVRI$ filters from the CFHT Megacam and CFH12K imagers \citep{balogh09}. 

In addition to the follow-up observations of the CNOC2 fields, group membership has also been redefined by \citet{wilman05} using more relaxed algorithm parameters than those used by \citet{carlberg01}.  The original search parameters were optimized such that the group-finding algorithm would identify dense, virialized groups, while the \citet{wilman05} catalog includes looser group populations that cover a wider range of dynamical states.  For this reason, the GEEC group catalog is ideal for the investigation of substructure within groups, since we are not restricted to dense group cores and are able to probe the surrounding large scale structure.

\subsection{Analysis of the GEEC Groups}
We apply the DS Test, as described in Section 2, to a subset of the GEEC groups.  Although there are roughly 200 groups in the GEEC catalog, the majority of systems have fewer than 10 members.  In Section 2.2.3, we determined that in order to obtain a reliable measure of substructure, the minimum number of member galaxies for the DS Test is $n_{\rm{members}} = 20$, which leaves us with a subset of 15 groups.  These groups are amongst the most massive GEEC groups with an average velocity dispersion of $\sim$525 km s$^{-1}$.

As previously mentioned, due to the relaxed membership allocation algorithm parameters, some of the GEEC groups have relatively large radial extents, and can be larger than the suggested maximum group virial radius ($r_{200}$) of 1.0 Mpc \citep{mamon07}, but for the purposes of detecting substructure we elect not to apply any radial cuts to our groups for our main analysis.  Our definition of substructure is liberal, in that we include structure that may be infalling or structure that is in nearby large scale structure but not necessarily bound to the host group.  Thus, to ensure that we do not eliminate any possible detection of substructure, we analyze all galaxies identified as group members by the FOF-algorithm applied in \citet{wilman05}.  Our decision not to apply radial cuts is further justified given that substructure is often in group and cluster outskirts \citep{wb90,zm98a}.

For the distribution parameters we estimate $\overline{\nu}^i_{\text{local}}$ and $\overline{\nu}$, from Equation \ref{delis}, as the group and local (i.e.\ $i$th galaxy and its $\sqrt{n_{\rm{members}}}$ nearest neighbours) canonical mean velocity, and $\sigma^i_{\text{local}}$ and $\sigma$, also from Equation \ref{delis}, as the local and group intrinsic velocity dispersion, computed following the method outlined in \citet{wilman05}.  The dispersion uncertainties are computed using the jackknife method \citep{efron82} and are given in Tables \ref{subgroups} and \ref{notsubgroups}.

\begin{table*}
\centering
\caption{Group properties and DS statistics for the GEEC groups with substructure \label{subgroups}}
\vspace{0.5cm}
\begin{tabular}{cccccccc}
\hline\hline
GEEC Group ID & $n_{\rm{members}}$ & $z_{group}$  & $\sigma$  & $\Delta/n_{\rm{members}}$ & $P$ value$^{a}$ & AD Test & Shape of Velocity\\
 & & & km s$^{-1}$ & &  & Classification &Dispersion Profile\\
\hline
25 & 28 & 0.362 & 491$\pm$32 & 0.693 &  0 & non-Gaussian & rising\\
208 & 22 & 0.269  & 530$\pm$45 & 0.841 &  0 & non-Gaussian & rising\\
226 & 86 & 0.359  & 944$\pm$17 & 1.37 & 0 & non-Gaussian & rising\\
320 & 29 & 0.245 & 463$\pm$33 & 0.751 & 0.00383 & Gaussian & rising\\
\hline
\end{tabular}
\newline
$^{a}$Using 100 000 `MC shuffles'.  We identify groups with $P$ values $<$ 0.01 as having significant substructure.
\end{table*}
\begin{table*}
\centering
\caption{Group properties and DS statistics values for the GEEC groups with no substructure \label{notsubgroups}}
\vspace{0.5cm}
\begin{tabular}{cccccccc}
\hline\hline
GEEC Group ID & $n_{\rm{members}}$ & $z_{group}$ & $\sigma$ & $\Delta/n_{\rm{members}}$ & $P$-value$^{a}$ & AD Test & Shape of Velocity\\
 & & & km s$^{-1}$  & & & Classification &Dispersion Profile\\
\hline
4 & 20 & 0.201  & 360$\pm$39 & 0.605 & 0.524 & non-Gaussian & flat\\
38 & 34 & 0.511 & 739$\pm$21 & 0.910 & 0.0372 & non-Gaussian & declining\\
104 & 27 & 0.145  & 365$\pm$25 & 0.659   & 0.0501 & Gaussian & flat\\
110 & 33 & 0.156  & 338$\pm$22 & 0.599  & 0.413 & Gaussian & declining\\
117 & 27 & 0.220  & 261$\pm$20 & 0.412  & 0.851 & Gaussian & flat\\
138 & 53 & 0.438  & 743$\pm$29 & 0.983 & 0.354 & non-Gaussian & flat\\
238 & 21 & 0.408 & 606$\pm$52 & 0.826  & 0.135 & non-Gaussian & flat\\
308 & 26 & 0.224  & 511$\pm$35 & 0.854  & 0.610 & Gaussian & declining\\
334 & 20 & 0.323  & 454$\pm$30 & 0.670  & 0.0563 & Gaussian & flat\\
346 & 32 & 0.373  & 439$\pm$20 & 0.612  & 0.232 & non-Gaussian & flat\\
362 & 24 & 0.460  & 666$\pm$43 & 0.861  & 0.0817 & Gaussian & rising\\
\hline
\end{tabular}
\newline
$^{a}$Using 100 000 `MC shuffles'.  We only identify groups with $P$ values $<$ 0.01 as having significant substructure.
\end{table*}

\begin{figure}
\centering
\includegraphics[width = 8cm, height = 8cm]{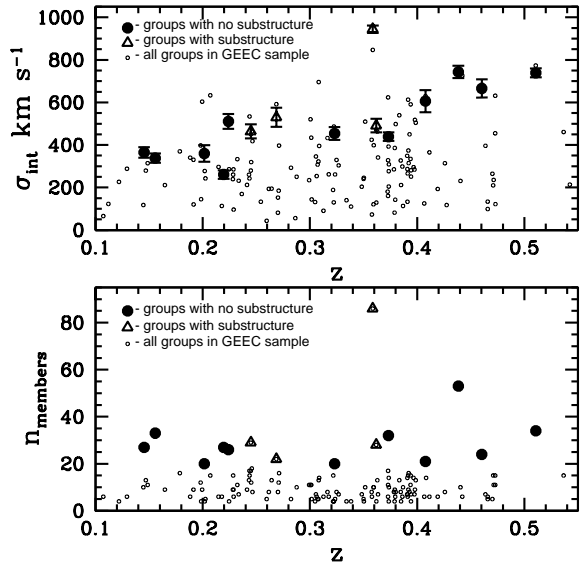}
\caption{Top:  Intrinsic velocity dispersion $\sigma_{int}$ versus the group redshift for the galaxy groups in our sample.  Solid circles indicate groups with no substructure and open triangles indicate groups with identified substructure according to the DS Test, at a 99 per cent confidence level.  The smaller open circles represent the dispersions of all the groups in the GEEC catalogue, the majority of which are not used in this analysis.  Our $n_{\rm{members}} \geq 20$ GEEC sample tend to have higher velocity dispersions.  Bottom:  $n_{\rm{members}}$ versus the group redshift for the galaxy groups in our sample.  Symbols are the same as the plot above.}
\label{sig_z}
\end{figure}

Of the 15 GEEC groups with $n_{\rm{members}} \geq 20$ we find that 4 groups, ($\sim$27 per cent) are identified as having substructure at the 99 per cent confidence level (c.l.).  In Table \ref{subgroups}, we list group properties, $\Delta/n_{\rm{members}}$-values, and $P$-values for the groups with substructure, identified as the systems that have $P$-values of less than 1 per cent.   In Table \ref{notsubgroups} we list the same values, but for groups without substructure, that is systems that have $P$-values greater than 1 per cent.  Also, it should be noted that although we list the $\Delta/n_{\rm{members}}$ critical values in Tables \ref{subgroups} and \ref{notsubgroups}, we rely on the P-values to identify substructure in our groups.  In addition, Tables \ref {subgroups} and \ref{notsubgroups} lists the dynamical properties of the groups to be discussed in Sections 4.1 and 4.2.  In Fig. \ref{sig_z}, we plot both the intrinsic velocity dispersion of the group (top) and the total number of group members (bottom) versus the mean group redshift.  Groups with and without substructure both span a wide range of velocity dispersions and group membership indicating that there is no apparent redshift bias with regards to group velocity dispersions or $n_{\rm{members}}$ for the DS Test.    

Although the minimum membership cut needed to determine a reliable percentage of groups that contain substructure is $n_{\rm{members}} = 20$, we can still apply the DS Test to systems with fewer members and establish a lower limit on this fraction.  Including groups with as few as ten members increases our sample size to 63 systems.  Applying the same methodology described above we find that 11 of our 63 groups ($\sim$17 per cent) are identified as having significant ($>$ 99 per cent c.l.) substructure.

\subsection{GEEC Groups with Substructure}
We examine the GEEC groups classified as having substructure by the DS Test in detail to determine if we can identify the regions of localized substructure.  By examining the available position and velocity information, we can find collections of galaxies that are kinematically distinct from the host group.

\begin{figure}
\centering
\includegraphics[width = 4.38cm, height = 4.38cm]{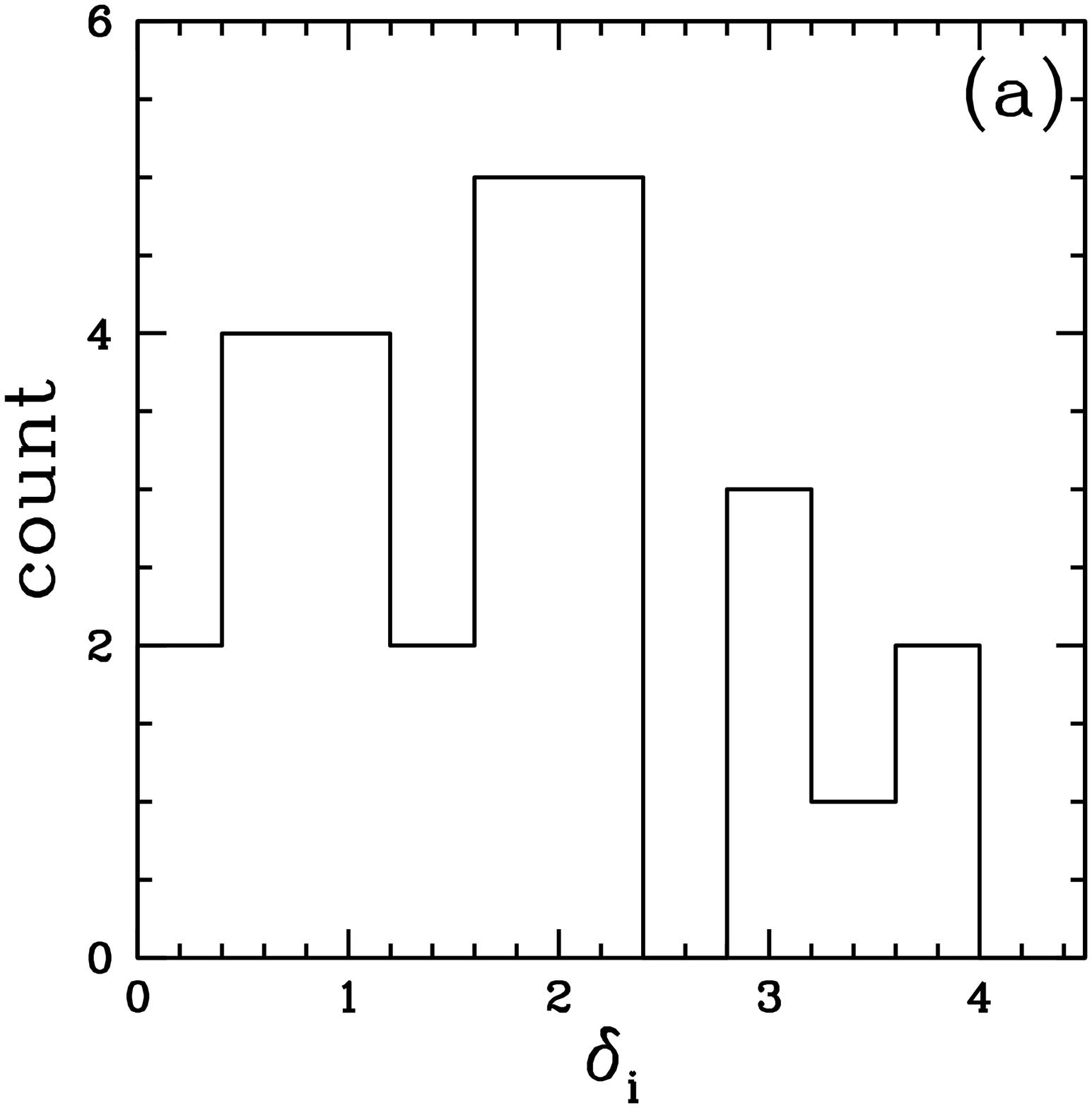}
\includegraphics[width = 4.38cm, height = 4.38cm]{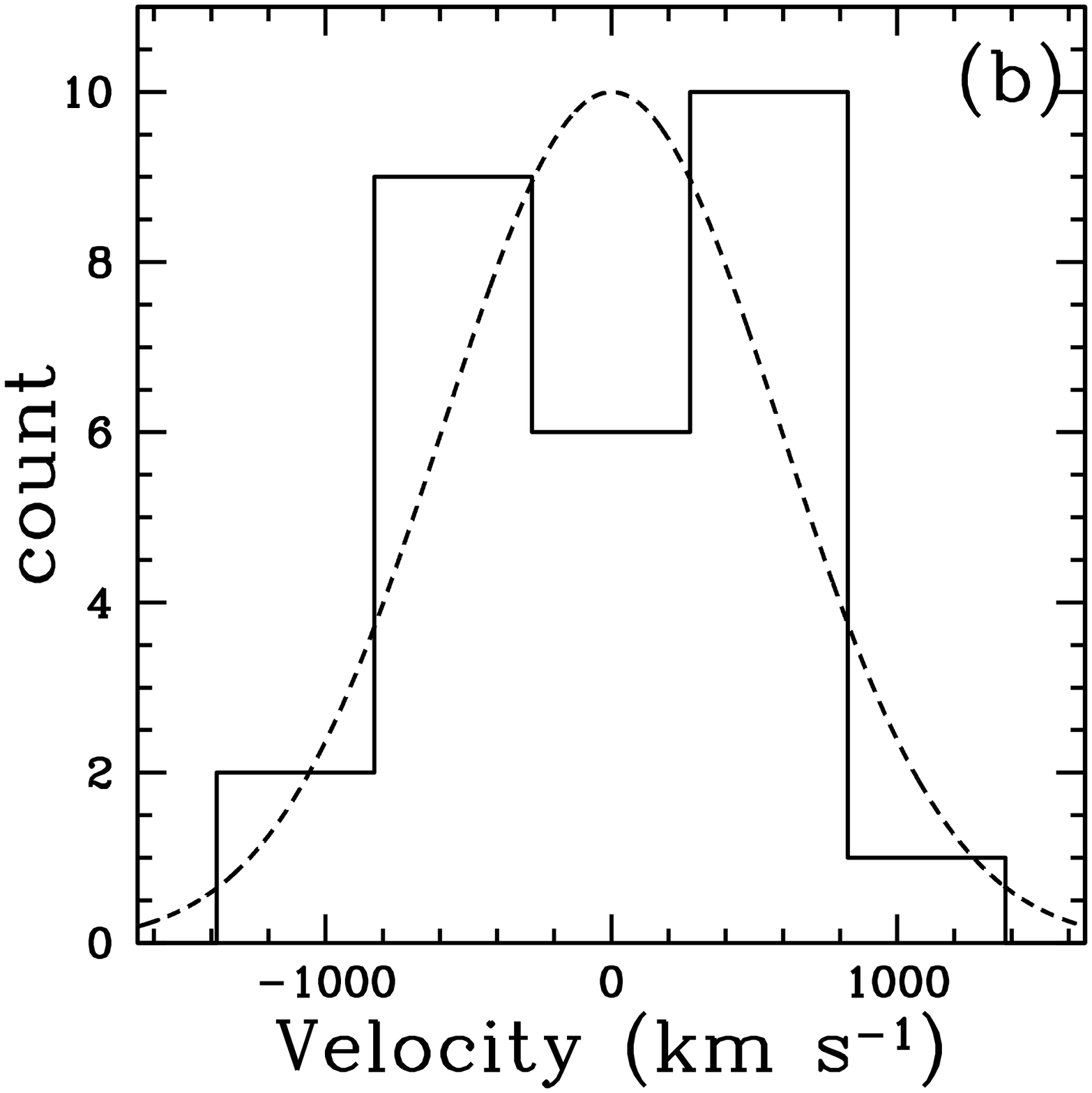}
\newline
\includegraphics[width = 4.38cm, height = 4.38cm]{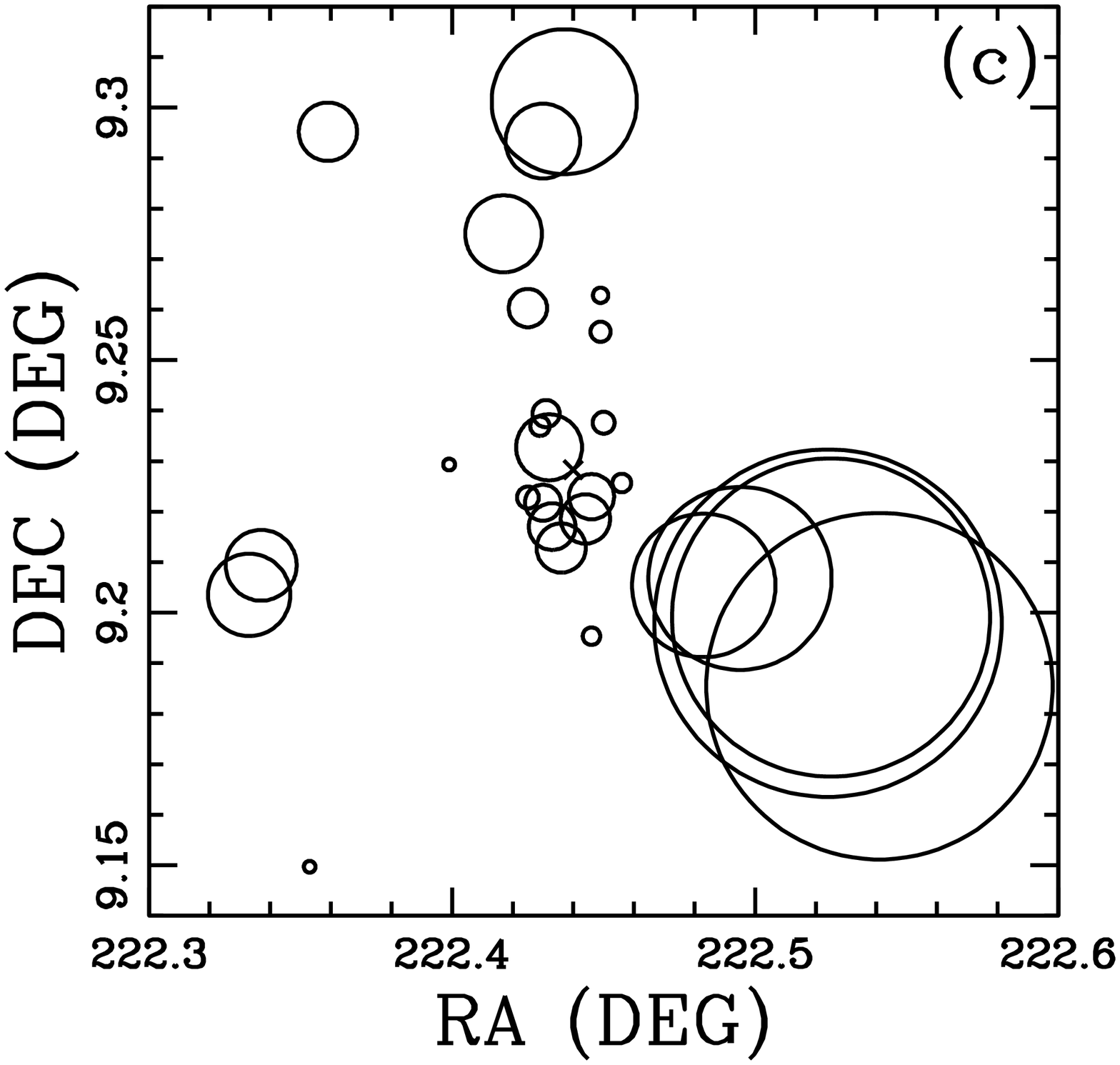}
\includegraphics[width = 4.38cm, height = 4.38cm]{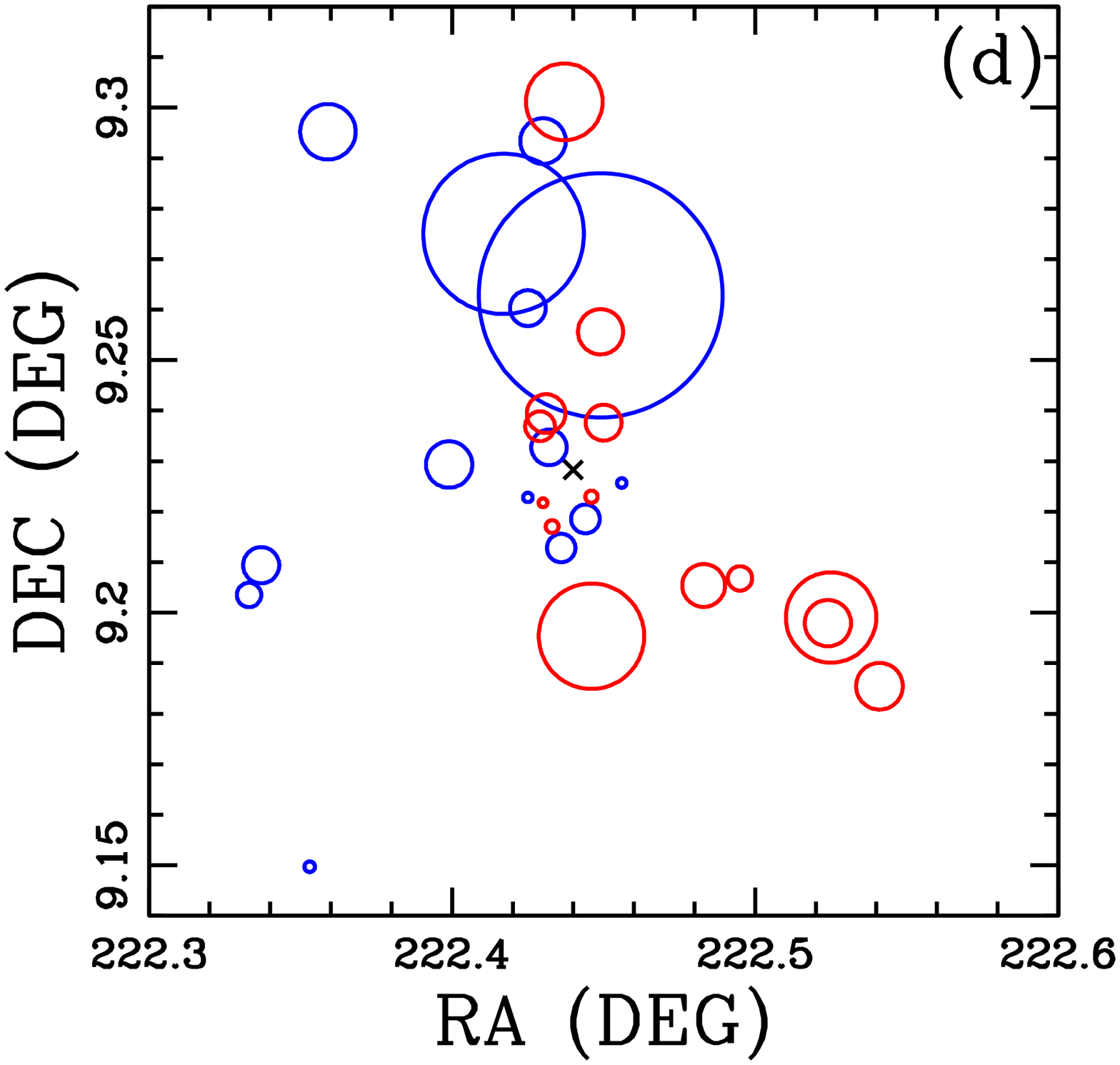}
\caption{GEEC Group 25.  (a): $\delta_{i}$ histogram. (b):  Histogram of the velocity distribution, where the dashed-line indicates the best fitting Gaussian velocity dispersion.  (c):  \citet{ds88} `bubble-plot' where the galaxy symbols scale with $\exp(\delta_{i})$.  (d): Position plot where the galaxy symbols scale with group-centric velocity (i.e.\  $\exp{\left(cz_{\rm{member}} - cz_{\rm{group}}\right)/350}$), blue symbols correspond to galaxies with negative group-centric velocities (i.e. $cz_{\rm{member}} - cz_{\rm{group}} < 0$) and red symbols correspond to galaxies with positive group-centric velocities  (i.e. $cz_{\rm{member}} - cz_{\rm{group}} > 0$).}
\label{g25}
\end{figure}

\begin{figure}
\centering
\includegraphics[width = 4.38cm, height = 4.38cm]{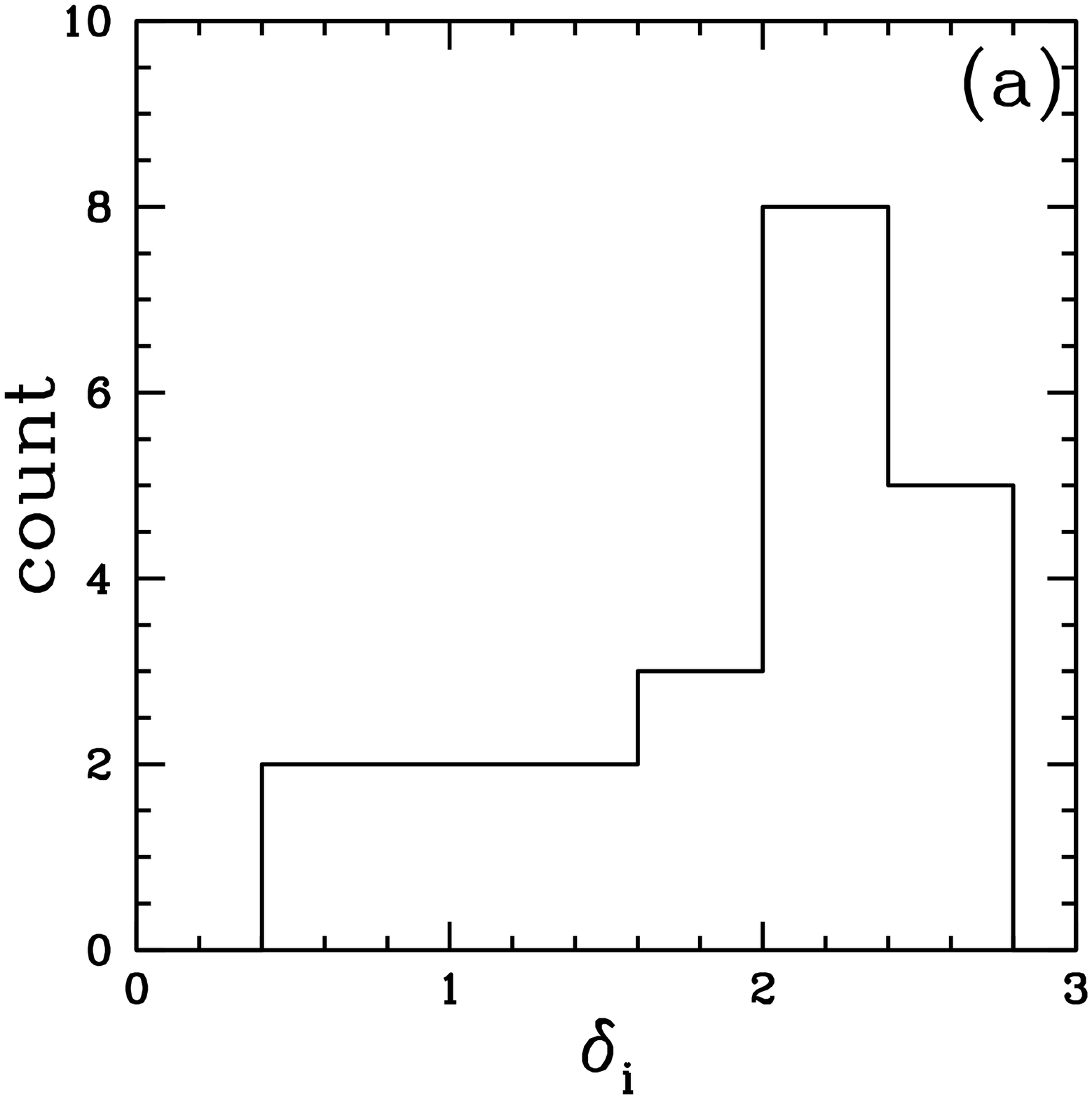}
\includegraphics[width = 4.38cm, height = 4.38cm]{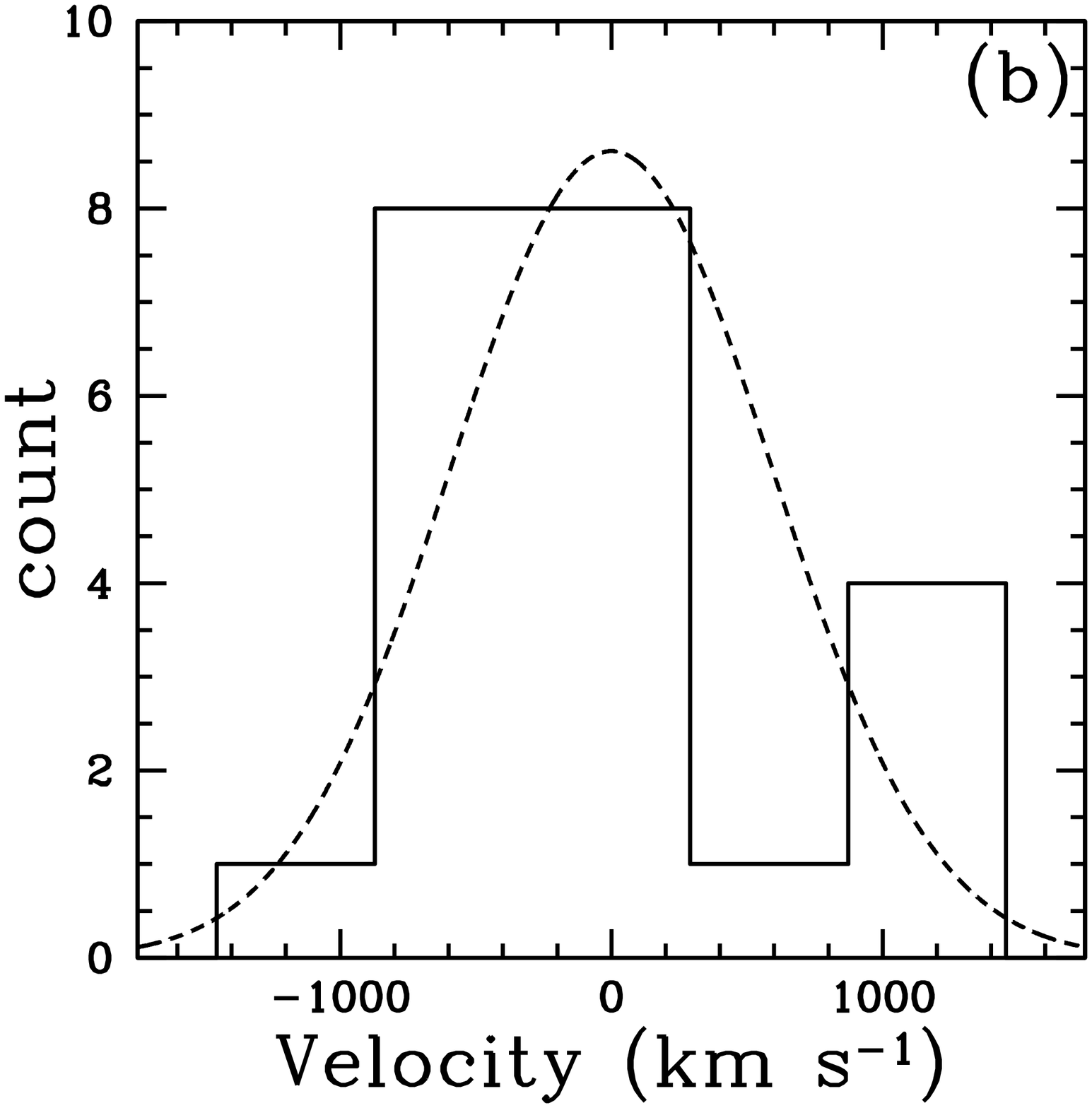}
\newline
\includegraphics[width = 4.38cm, height = 4.38cm]{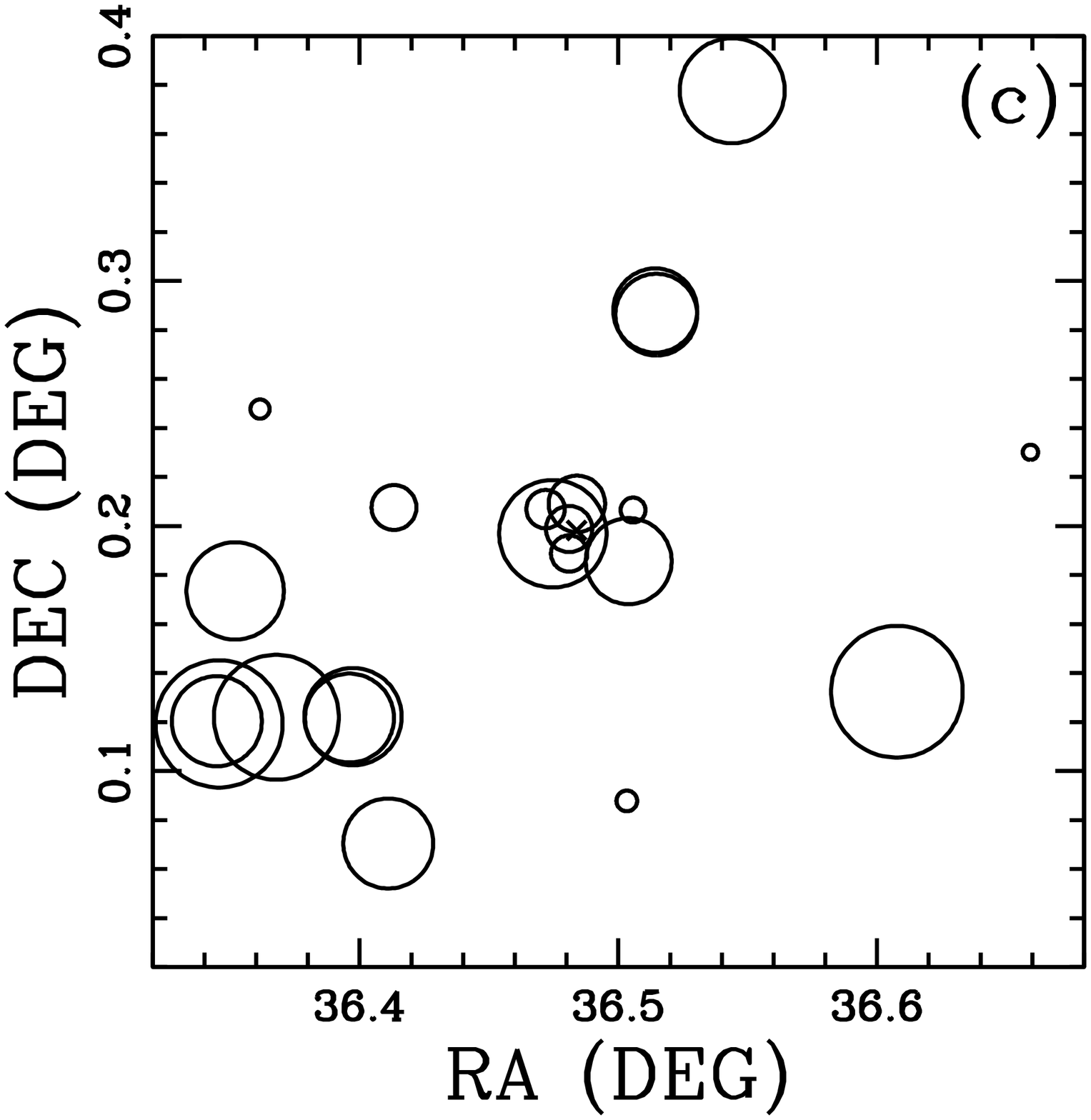}
\includegraphics[width = 4.38cm, height = 4.38cm]{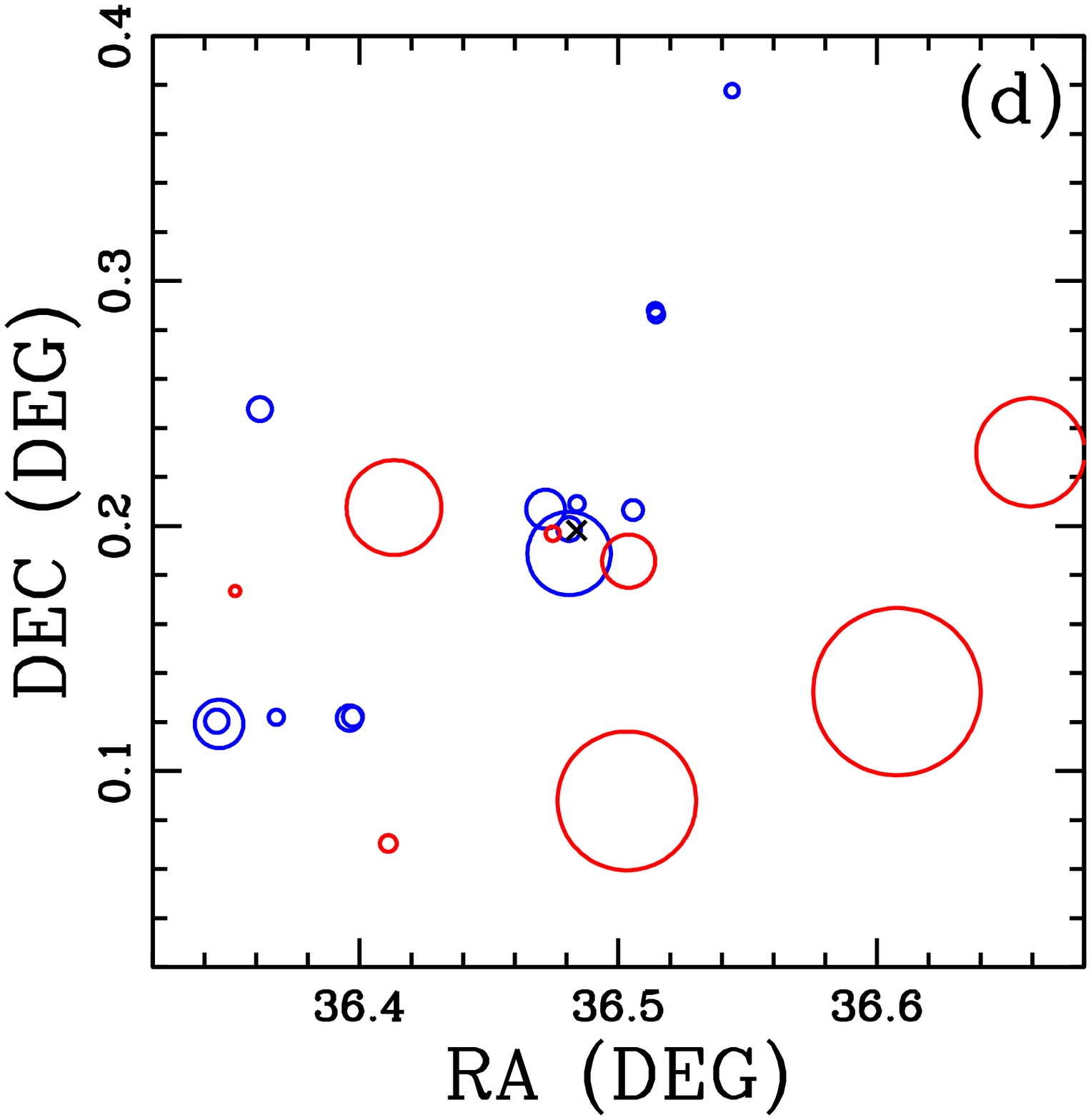}
\caption{Same as Fig. \ref{g25} but for GEEC Group 208 and the velocity weighted plot now scales as $\exp{\left(cz_{\rm{member}} - cz_{\rm{group}}\right)/400}$.}
\label{g208}
\end{figure}

\begin{figure}
\centering
\includegraphics[width = 4.38cm, height = 4.38cm]{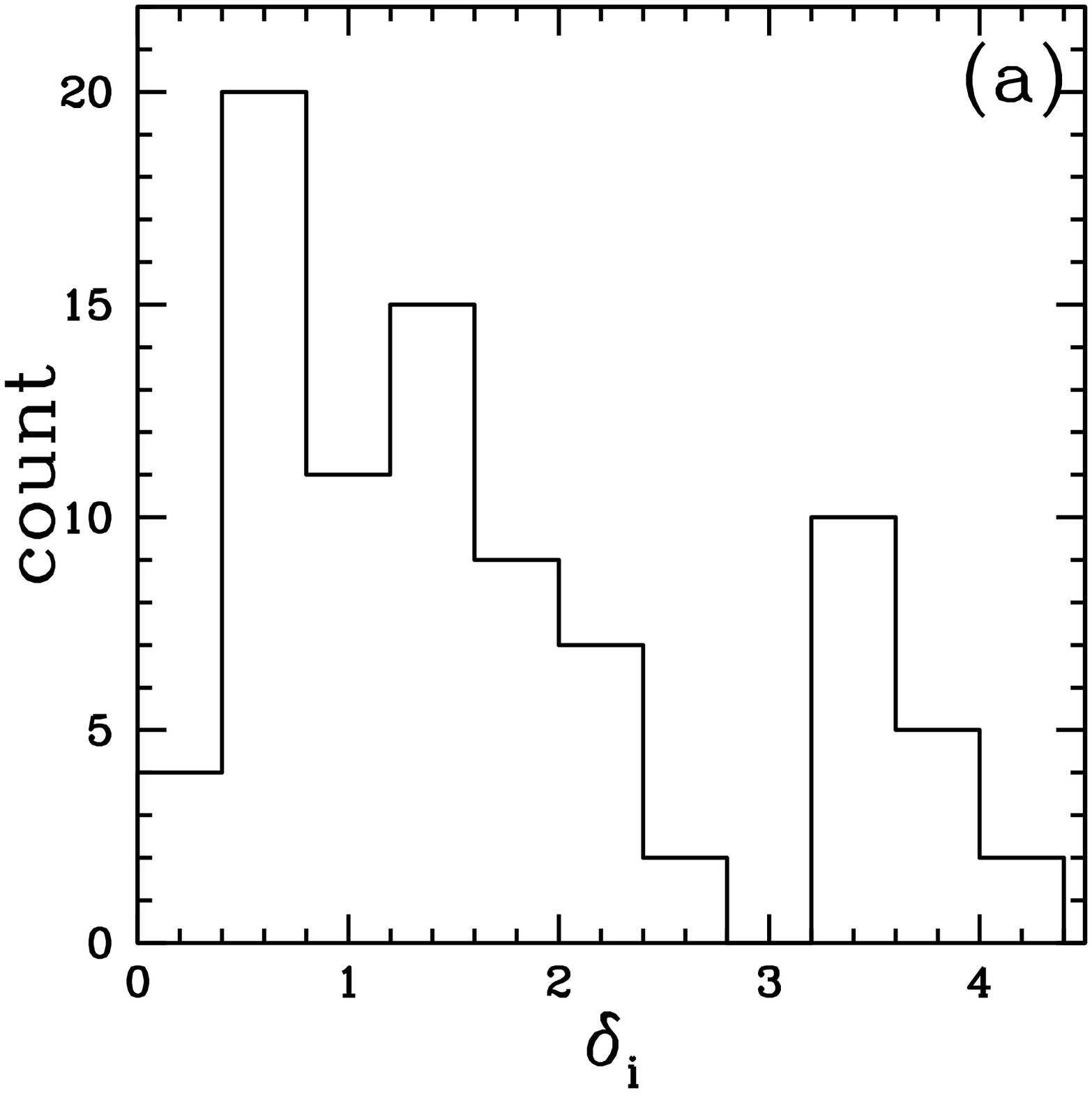}
\includegraphics[width = 4.38cm, height = 4.38cm]{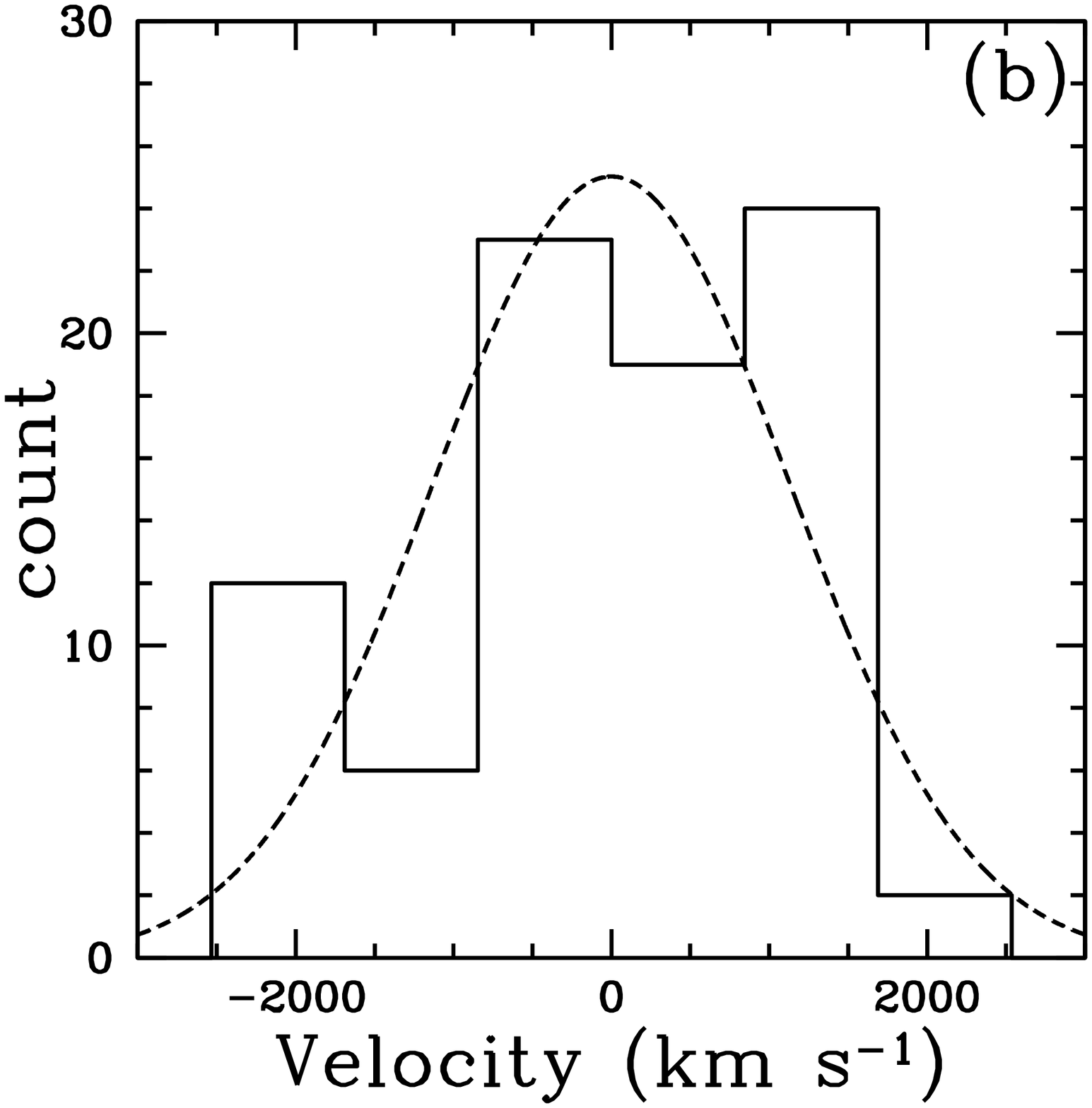}
\newline
\includegraphics[width = 4.38cm, height = 4.38cm]{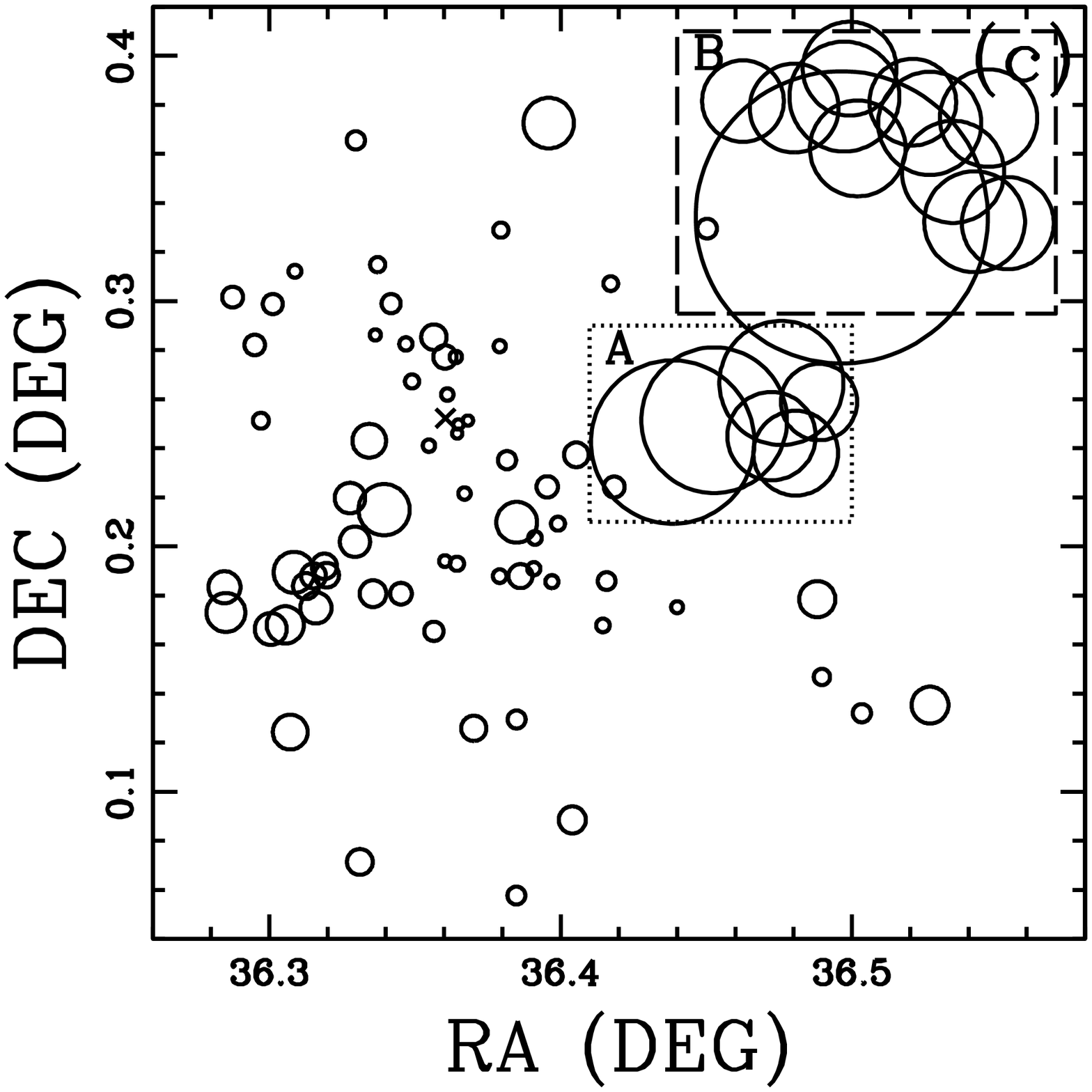}
\includegraphics[width = 4.38cm, height = 4.38cm]{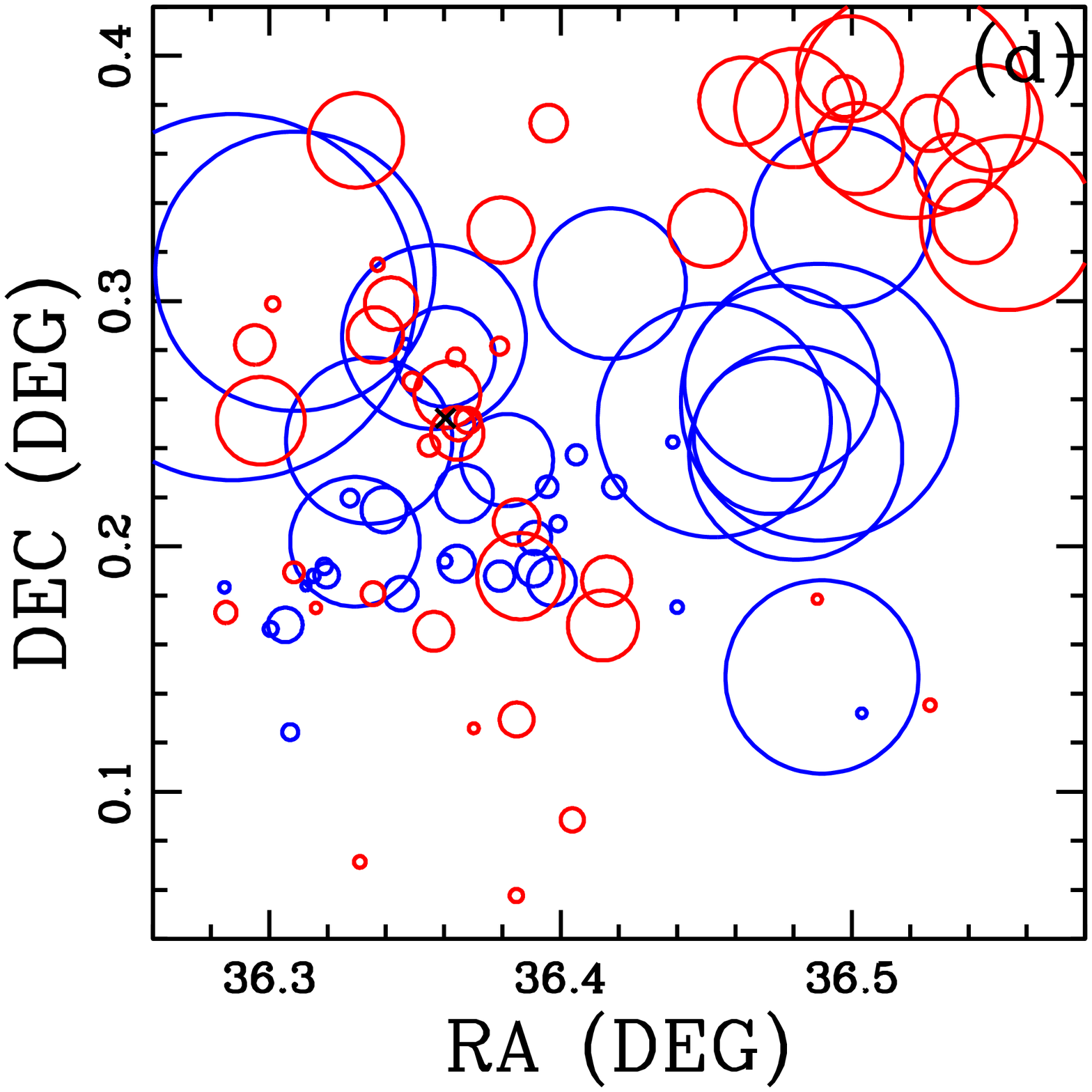}
\caption{Same as Fig. \ref{g25} but for GEEC Group 226 and the velocity weighted plot now scales as $\exp{\left(cz_{\rm{member}} - cz_{\rm{group}}\right)/600}$.  The dotted box encompasses the first identified region of local substructure (region A) and the long dashed box encompasses the second region of local substructure (region B).}
\label{g226}
\end{figure}

\begin{figure}
\centering
\includegraphics[width = 4.38cm, height = 4.38cm]{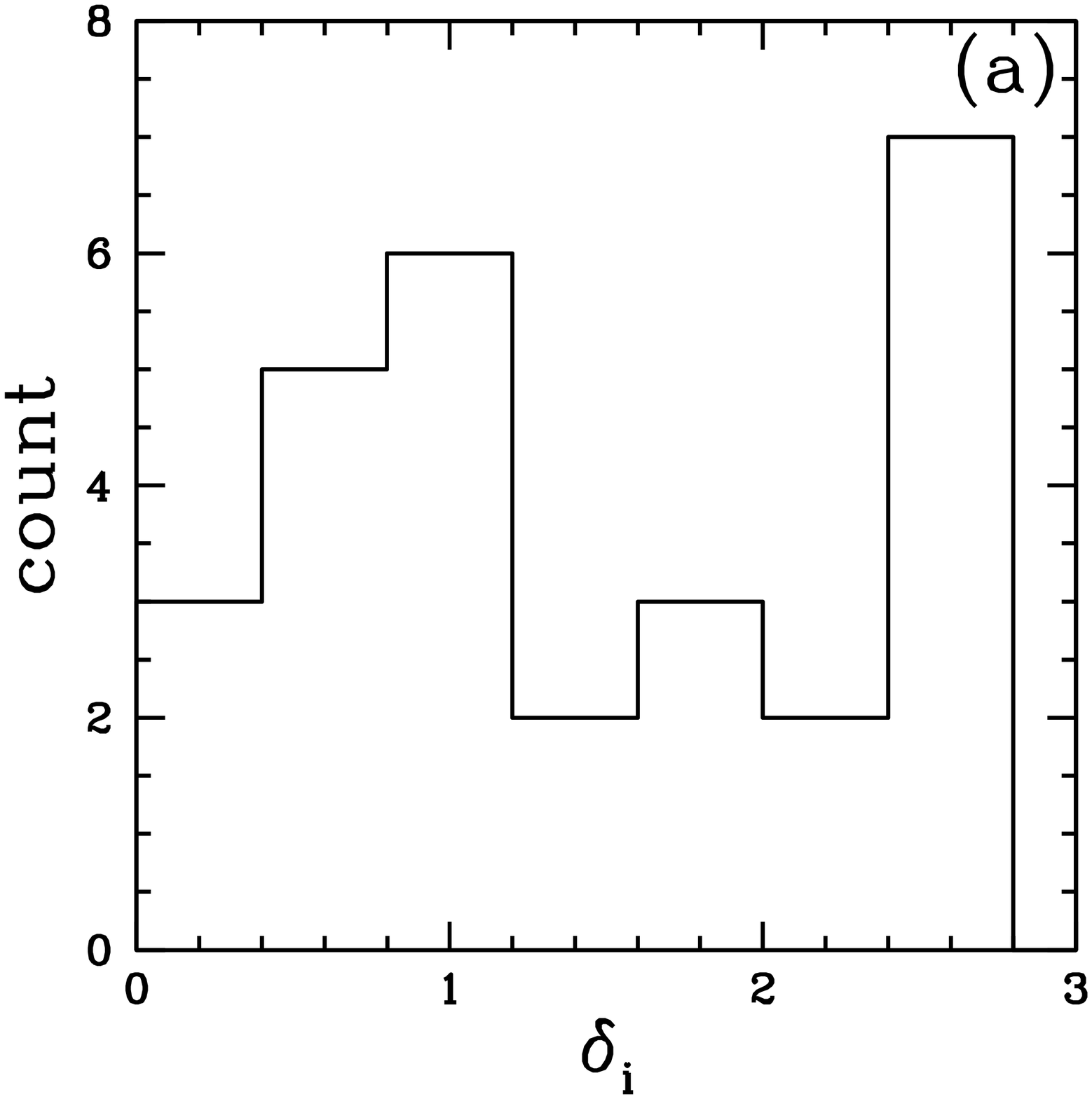}
\includegraphics[width = 4.38cm, height = 4.38cm]{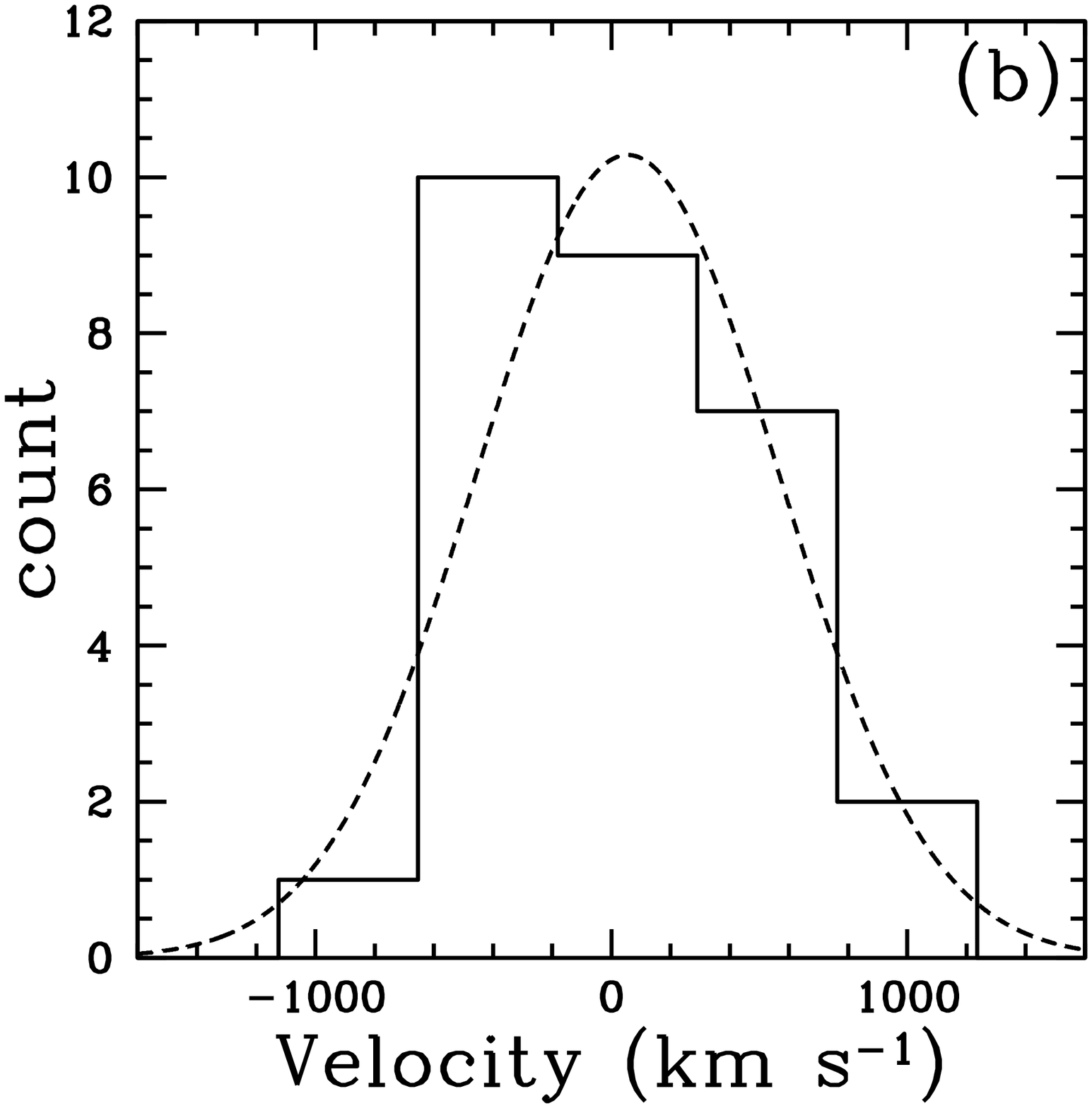}
\newline
\includegraphics[width = 4.38cm, height = 4.38cm]{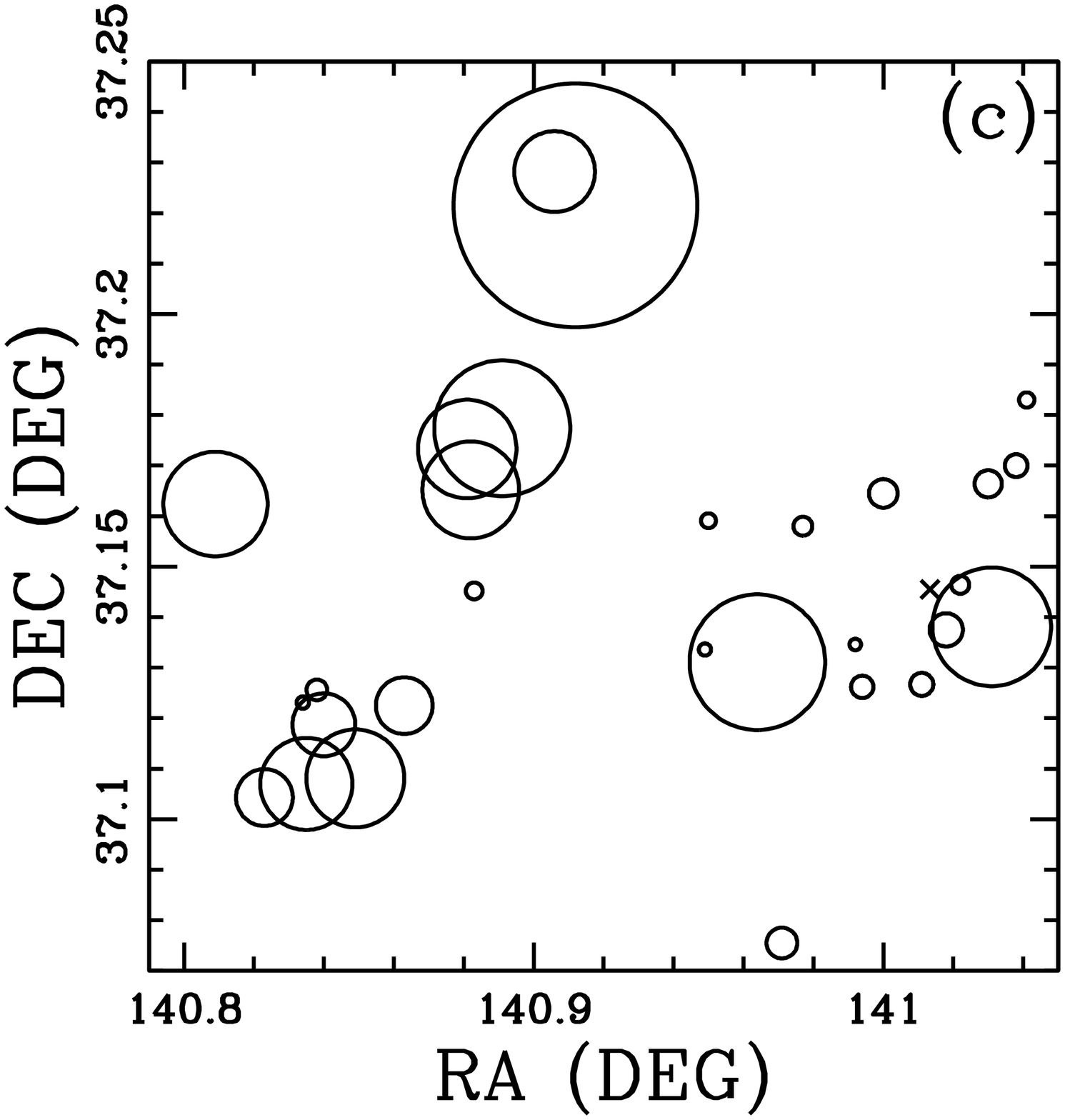}
\includegraphics[width = 4.38cm, height = 4.38cm]{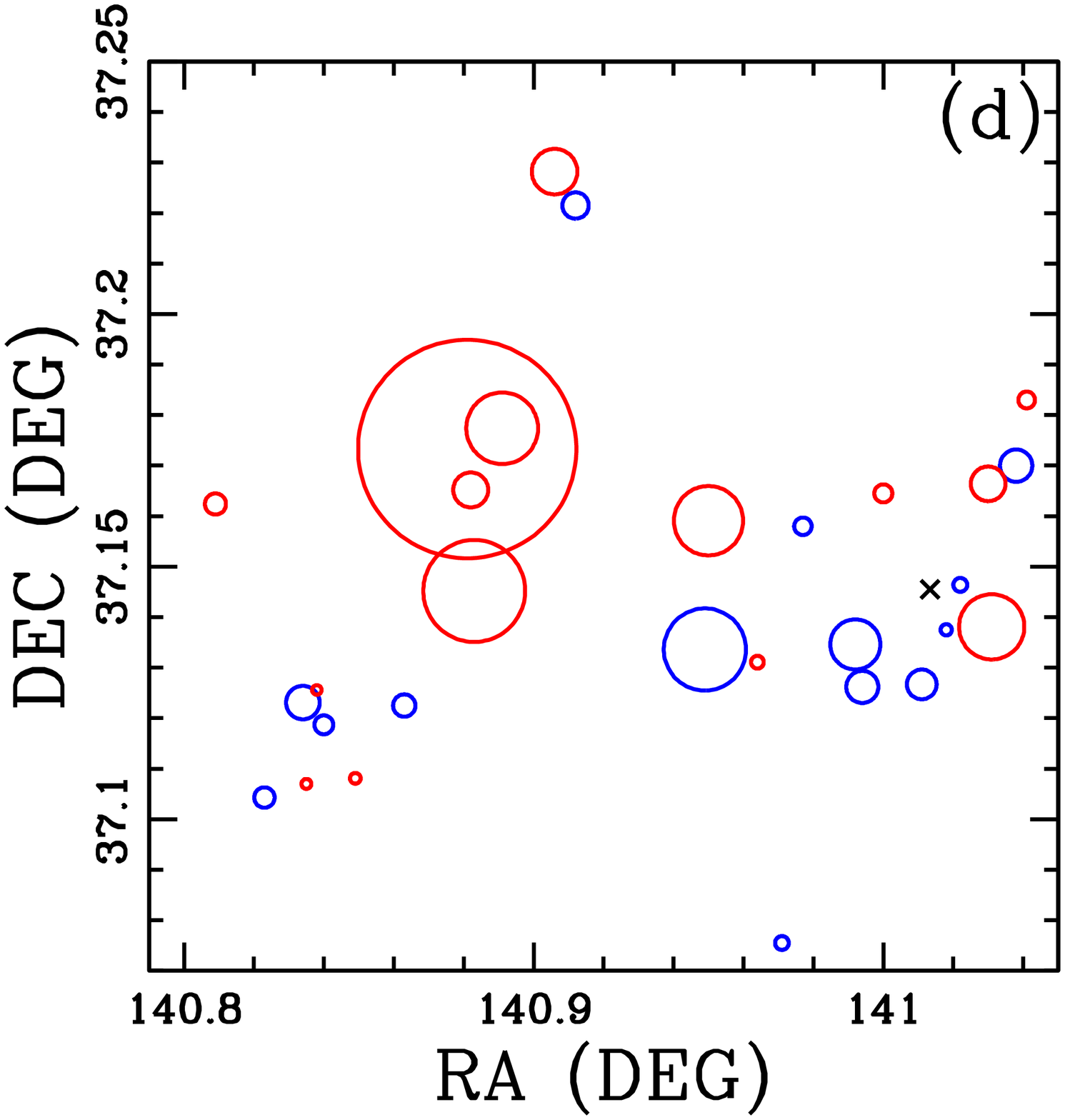}
\caption{Same as Fig. \ref{g25} but for GEEC Group 320.}
\label{g320}
\end{figure}

In the following analysis, we focus on small subgroups of galaxies that may be part of some localized substructure.  This is done by looking at the $\delta_{i}$ histogram (Figs. \ref{g25}(a) - \ref{g320}(a)), velocity histograms (Figs. \ref{g25}(b) - \ref{g320}(b)), `bubble-plots' (i.e.\ position plots of the group members weighted by $\exp{\delta_{i}}$ \citep{ds88}  (Figs. \ref{g25}(c) - \ref{g320}(c))  and group-centric radial velocity (i.e.\ $cz_{\rm{member}} - cz_{\rm{group}}$) weighted position plots (Figs. \ref{g25}(d) - \ref{g320}(d)).

The $\delta_{i}$ and velocity histograms provide an estimate on the amount of substructure and the dynamical state of the groups.  The $\delta_{i}$ histogram gives an overall view of the kinematic deviations and the velocity histogram can be used to identify non-Gaussian features, such as multiple peaks.  In order to look for local regions of substructure, one must look at both the `bubble-plot' and velocity weighted position plots in Figs. \ref{g25} - \ref{g320}.  The `bubble-plots' allow for the visual identification of candidate regions of local substructure.  Since the size of the symbols in the `bubble-plot' scales with a galaxy's $\delta_{i}$ value, large symbols correspond to strong local kinematic deviations from the global values.  Thus, a collection of neighbouring galaxies with similarly large symbols, such as region A in Fig. \ref{g226}(c), could indicate a kinematically distinct system.  In order to confirm that these candidate regions are truly distinct, one must check that the candidate substructure galaxies also have similar velocities, since the sign or direction of the galaxy velocity is not taken into account in the DS Test (Equation \ref{delis}).  To determine this, we look at the group-centric velocity weighted position plot to see if the candidate local substructure galaxies have similar velocities.  In Figs. \ref{g25}(d) to \ref{g320}(d) the red symbols correspond to positive group-centric velocities (i.e.\ $cz_{\rm{member}} - cz_{\rm{group}} > 0$), blue symbols correspond to negative group-centric velocities (i.e.\ $cz_{\rm{member}} - cz_{\rm{group}} < 0$) and the symbol size scales with the magnitude of the velocity offset.  We only identify galaxies as part of local substructure if neighbouring galaxies have similar large kinematic deviations, shown in the `bubble-plots' \emph{and} similar group-centric radial velocities, shown in the weighted position plots. 

We will now discuss each of the four GEEC Groups with substructure in detail.  Using the methodology described above, we search for candidate local regions of substructure in our sample.  GEEC Group 25 (Fig. \ref{g25}) has a collection of five galaxies, just south-east of the group centre, that all have high $\delta_{i}$ values and comparable velocities.  Similarly in GEEC Group 208 there are seven galaxies that lie south-west of the group centre, with equally high $\delta_{i}$ values (Fig. \ref{g208}(c)).  Though, when we look at these same galaxies in the velocity weighted position plot (Fig. \ref{g208}(d)), only five of these seven galaxies have comparable radial velocities.  This result highlights the importance of looking at both the `bubble-plot' and velocity weighted positions plots, as galaxies with large kinematic deviations may be correlated in position-space, but not in velocity-space. 

In Fig. \ref{g226}, we show the substructure analysis plots for GEEC Group 226 and from the `bubble-plot' we see that there are two possibly distinct regions of localized substructure.  The first region lies directly north-east of the group centre (region A in Fig. \ref{g226}(c)) and includes the galaxy with the highest $\delta_{i}$ value.  All of the members within this given substructure have similar radial velocities (Fig. \ref{g226}(d)).  The second region of interest contains eleven galaxies that lie in the very north-east corner of the `bubble-plot' (region B in Fig. \ref{g226}(c)).  Again, all of the galaxies within this particular substructure have similar group-centric velocities, though in the opposite direction of the members in region A. 

\clearpage
From the `bubble-plot' of GEEC Group 320 (Fig. \ref{g320}(c)), we see two possible regions of substructure; one just north-west of centre and another further south-west of centre.  The structure near the group centre does not appear to be very localized, as the velocities, though in the same direction, have significantly different magnitudes (Fig. \ref{g320}(d)).  The second region of high $\delta_{i}$ values may actually be two separate structures.  The velocity weighted position plot in Fig. \ref{g320}(d) reveals that three of the galaxies in the structure have similar negative group-centric velocities (i.e. $cz_{\rm{member}} - cz_{\rm{group}} < 0$) and two have similar positive velocities (i.e. $cz_{\rm{member}} - cz_{\rm{group}} > 0$).  Therefore, we identify the collection of three galaxies, with the negative group-centric radial velocities, as the best candidate of substructure within this system.

In each of the GEEC groups with substructure, it appears that our identified regions of localized substructure lie on the outskirts or edges of the group, and may either be infalling onto a pre-existing system or in the nearby large scale structure, but not necessarily bound or infalling.  This result is in agreement with similar studies of substructure in local groups \citep{zm98a} and clusters \citep{wb90}.

It should also be noted that although the DS Test has been shown to be reliable for group-sized systems, the number of members within the substructure can affect the results of the statistic.  As discussed in Section 2.2, and shown in more detail in the Appendix, fewer(more) true members of substructure can increase(decrease) the rate of false negatives.  However, we also show that the rate of false positives is consistently low for all of our mock groups (see Section 2.2).  Thus, any detection of substructure in these systems is likely to be real.

\subsection{Is the localized substructure gravitationally bound to the group?}
\begin{figure}
\centering
\includegraphics[width = 4cm, height = 4cm]{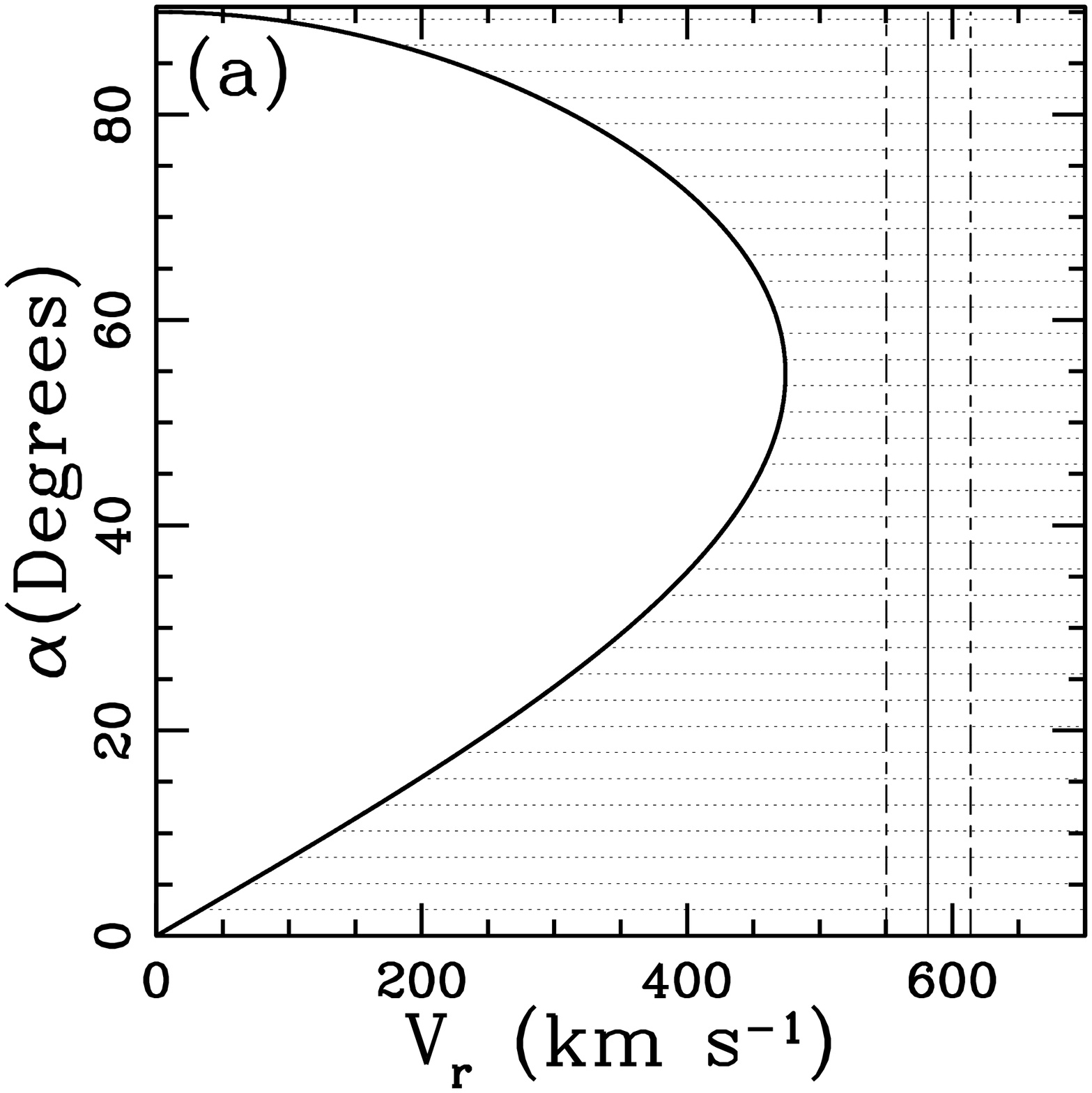}
\includegraphics[width = 4cm, height = 4cm]{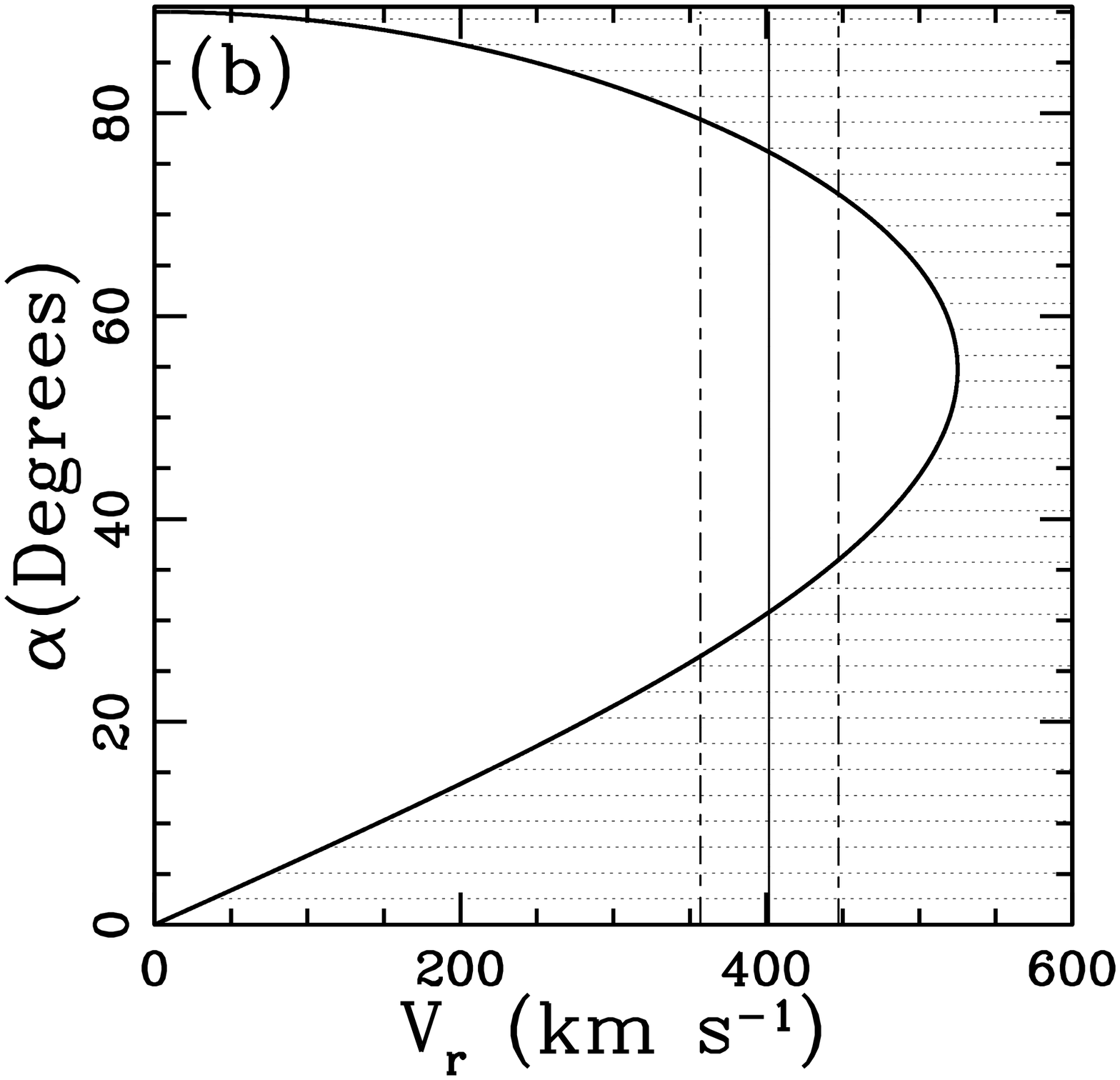}
\newline
\includegraphics[width = 4cm, height = 4cm]{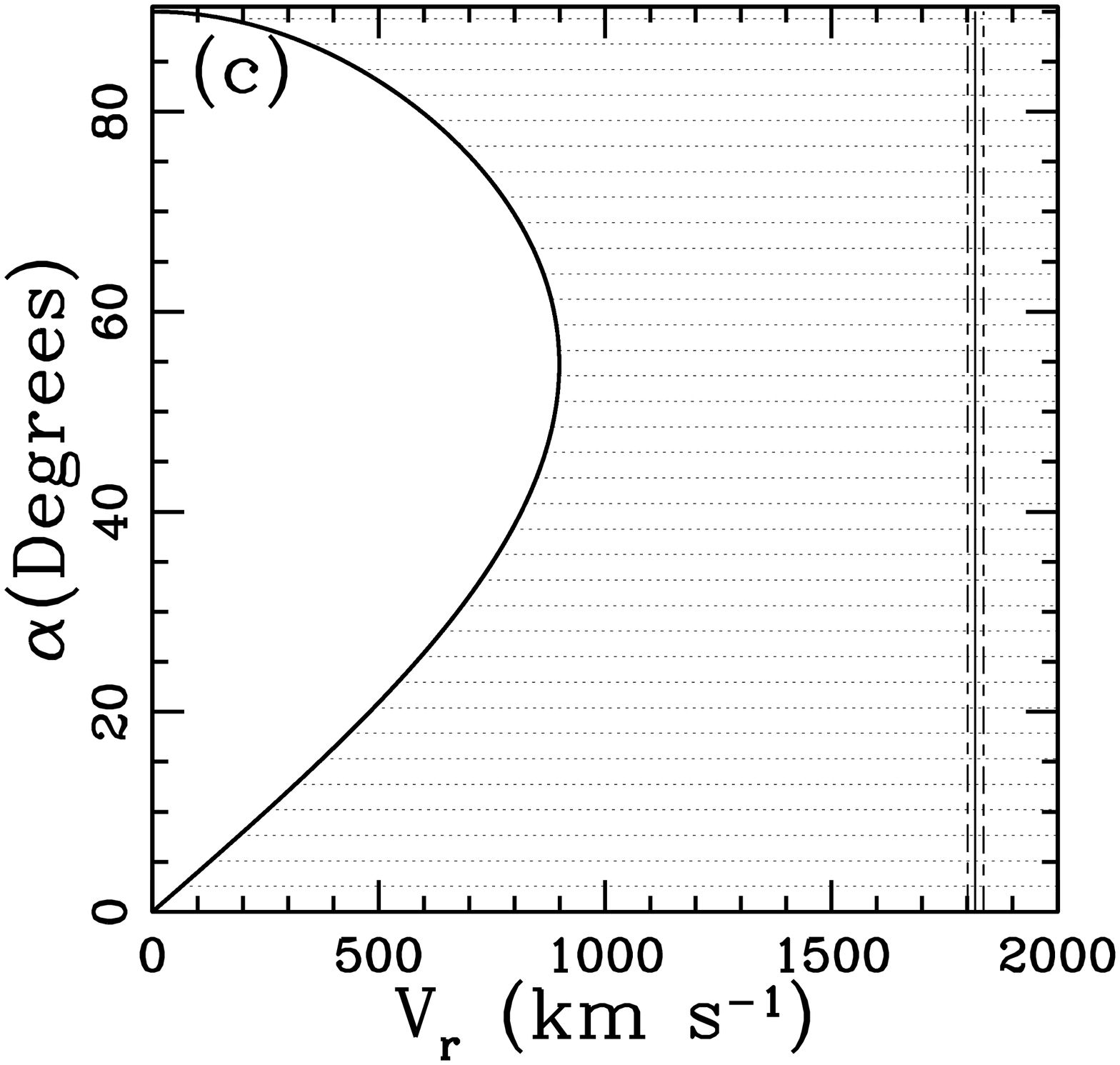}
\includegraphics[width = 4cm, height = 4cm]{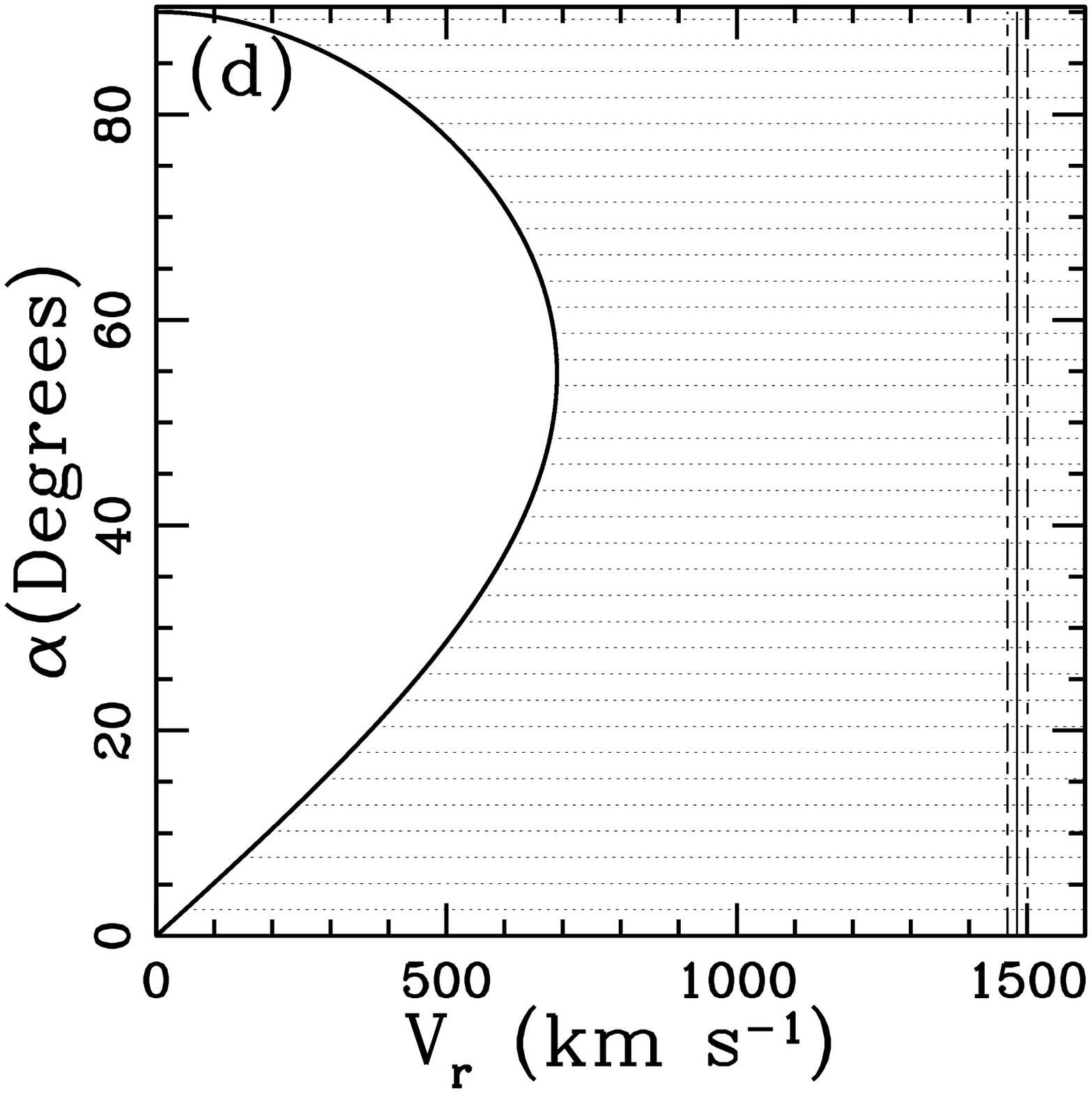}
\newline
\includegraphics[width = 4cm, height = 4cm]{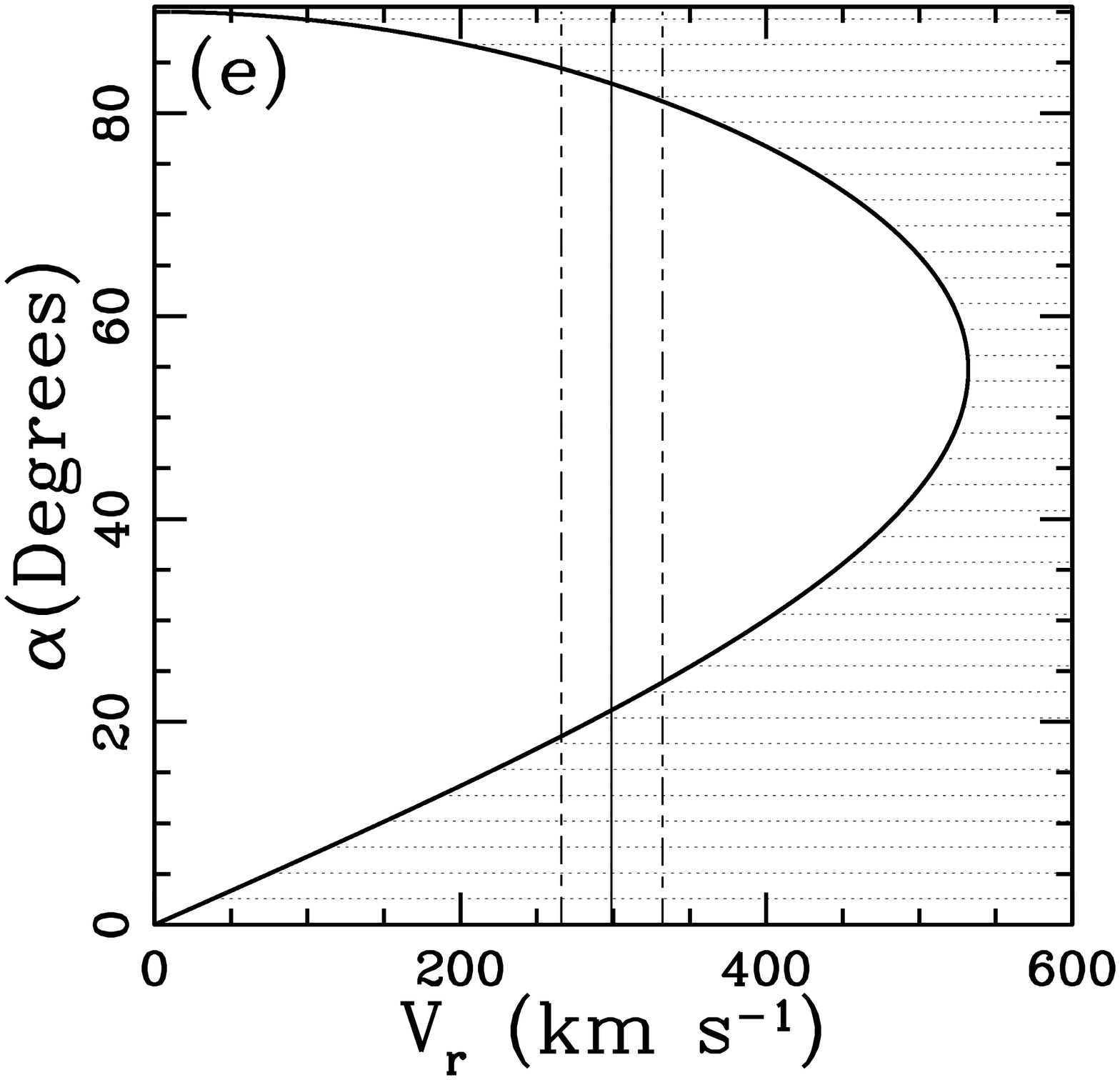}
\caption{$\alpha-V_{r}$ plot for GEEC Groups (a) 25, (b) 208, (c) region A and the host group of 226, (d) region B and the host group of 226 (d) and (e) 320 .  The unshaded regions correspond to bound solutions and shaded regions correspond to unbound solutions of the virial theorem given by Equation \ref{vt}.  The solid line corresponds to the measured value of $V_{r}$ (i.e.\ the line-of-sight velocity difference between the substructure and host group centres).  The dashed lines correspond to one-sigma deviations, taken to be the errors on the intrinsic velocity dispersion.}
\label{vralpha}
\end{figure}   

\indent{}The local regions of substructure we detect lie on the group outskirts and are possibly bound and infalling, or not bound, but close in large scale structure.  In order to help distinguish between these two possibilities, we apply a simple bound test to estimate whether or not the regions of localized substructure are gravitationally bound to the host group.  This is done by computing the limits for bound systems, using a variation of the virial theorem as discussed in \citet{beers82}, and determining if the substructure falls within these limits.  \citet{beers82} state that for a two-body system on a linear orbit, the Newtonian limit for gravitational binding, projected onto the sky is given by

\begin{equation}
\centering
V_{r}^{2}R_{p} \leq 2GM \sin^{2} \alpha \cos \alpha,
\label{vt}
\end{equation}
where,
\begin{equation}
\centering
V_{r} = V \sin \alpha, \hspace{0.5cm} R_{p} = R \cos \alpha,
\label{VrRp}
\end{equation}
\\
\noindent and where $\alpha$ is the angle between the line joining the two-body system and the plane of the sky \citep[see fig. 7 of][]{beers82}, $M$ is the \emph{total} mass of the entire system (substructure plus host group), and $R$ and $V$ are the true (3-dimensional) positional and velocity separations between the two objects.  $V_{r}$ is the line-of-sight relative velocity between the two bodies and $R_{p}$ is the projected separation, both of which are measurable quantities.  

The unknown quantity in Equation \ref{vt} is the projection angle $\alpha$.  Thus, to compute the limit between bound and unbound systems, one must work in $\alpha-V_{r}$ space to determine the probability that the system is gravitationally bound for any given projection angle.  This is achieved by setting $0^{\circ} \leq \alpha \leq 90^{\circ}$ and solving for $V_{r}$ in Equation \ref{vt}, producing a distinct line in $\alpha-V_{r}$ space that clearly separates bound and unbound solutions (Fig. \ref{vralpha}).  One can then compute the probability of a bound solution, for a given projection angle, using the $\alpha-V_{r}$ plot.   

To apply this methodology to our group sample, we treat our identified regions of local substructure as one-body and the remaining galaxies as the second-body, which we refer to as the `host' group.  The total mass of the system is taken to be the virial mass, $M_{200}$ (i.e. the total mass within a radius that encloses a mean density of 200 times the critical density of the Universe at the redshift of the galaxy), of the GEEC Groups as computed in \citet{balogh09} and given by
 \begin{equation}
 \centering
 M_{200} = \frac{3^{3/2}\sigma^{3}}{G}\frac{1}{10H_{0}\left(1+ z\right)^{1.5}},
 \label{m200}
 \end{equation}
 \\
 where $\sigma$ is the measured intrinsic velocity dispersion.  The measured $V_{r}$ and $R_{p}$ values are taken to be the distance between the $R$-band luminosity-weighted centres of the local substructure and the `host' group centre, along the line-of-sight ($V_{r}$) and projected on the sky ($R_{p}$).

In Fig. \ref{vralpha}, we plot the $\alpha-V_{r}$ plots for the five candidate regions of local substructure, as discussed in Section 3.3.  The shaded regions indicate the areas spanned by unbound solutions, as given by Equation \ref{vt}, the solid black vertical line is the measured value of $V_{r}$, and the long-short dashed vertical lines indicate one sigma deviations, taken to be the intrinsic velocity dispersion error.  

Using the methodology described above and Fig. \ref{vralpha}, we conclude that the identified local substructures in GEEC Groups 25 and 226 are not bound to the host group, while for GEEC Groups 208 and 320 the detected substructure is likely bound to the host group.

\subsection{Substructure within 1 Mpc of the Group Centroid}

\indent{}In the previous section we analyzed a subset of 15 GEEC groups, with $n_{\rm{members}} \geq 20$, without applying any radial cuts.  Here we apply a 1.0 Mpc cut, which is the suggested maximum virial radius for groups \citep{mamon07}, on the same subset of groups and re-apply the DS Test.  Applying this radial cut, while still requiring a minimum number of 20 member galaxies, reduces our sample from 15 to 5 groups (GEEC Groups 110, 138, 226, 308 and 346).  With this radial cut, we find that all 5 groups are identified as not having substructure (see Table \ref{DSradcut}) according to the DS Test.  The only group that was previously identified as having substructure, prior to the 1.0 Mpc cut, is GEEC Group 226.  In Fig. \ref{g226}, it is evident that although there are galaxies with relatively high $\delta_{i}$ values close to the group centre, the members with the highest $\delta_{i}$ values lie near the edge of the group.  In fact the most significant feature in the `bubble-plot' (Fig. \ref{g226}) is on the top-right corner of the plot, far from the group centre.  

If we include all of the groups with $n_{\rm{members}} \geq 10$ after a 1.0 Mpc radial cut, we find that 2 out 33 groups ($\sim$6 per cent) of our sample contains significant substructure.  Again, we note that for systems with fewer than 20 members, the DS Test can only provide a lower limit on the amount of substructure present.

\begin{table}
\caption{Groups properties and DS statistics values for the GEEC Groups with a 1.0 Mpc radius cut \label{DSradcut}}
\vspace{0.5cm}
\begin{tabular}{ccccc}
\hline\hline
GEEC Group ID & $n_{\rm{members}}^{a}$  & $\overline{\nu}$ & $\sigma_{int}$ & $P$-value\\
 & & km s$^{-1}$ & km s$^{-1}$ & \\
\hline
110 & 26 & -15.3 & 350 & 0.403\\
138 & 23 & -226 & 730 & 0.792\\
226 & 25 & -174 & 847 & 0.110\\
308 & 25 & 2.52 & 512 & 0.507\\
346 & 26 & -80.4 & 434 & 0.128\\
\hline
\end{tabular}
\newline
$^{a}$Group membership after the 1.0 Mpc radius cut
\end{table}

\section{Correlations between Substructure and Other Indicators of Dynamical State}
Having identified the GEEC groups that contain substructure, we can also look at other dynamical properties, i.e.\  the velocity distributions and velocity dispersion profiles (VDPs), to determine if there are any correlations with substructure.  

\subsection{Comparison with the Dynamical Classification of Velocity Distribution}
\indent{}If a correlation does exist between substructure and recent galaxy accretion, one would expect that the groups with substructure should also be dynamically complex, with perhaps non-Gaussian velocity distributions.  In \citet{hou09}, we established a classification scheme to distinguish between dynamically relaxed and complex groups.  Using the Anderson-Darling (AD) goodness-of-fit test, we are able to determine how much a system's velocity distribution deviates from Gaussian.  This is done by comparing the cumulative distribution function (CDF) of the ordered data, which in our case is the observed velocity distribution, to the model Gaussian empirical distribution function (EDF) using computing formulas given in \citet{D'agostino}.  The AD statistic is then converted into a probability, or $P$-value, using results determined via Monte Carlo methods in \citet{nelson98}.  A system is then considered to have a non-Gaussian velocity distribution if its computed $P$-value is less then 0.01 corresponding to a 99 per cent c.l.

%

We now apply this scheme to our sample of 15 GEEC groups, with $n_{\rm{members}} \geq 20$ and no radial cut, to compare the dynamical state with the detection of substructure\footnote{The dynamical classifications in this paper differ from those in \citet{hou09}.  The reason for this difference is that in \citet{hou09} we applied a 1.0 Mpc radius cut to the GEEC groups, whereas no radial cut is applied in this analysis.}.  The results of our dynamical classification scheme indicate that 8 of the 15 groups are classified as having non-Gaussian velocity distributions, at the 99 per cent c.l., and are thus dynamically complex (see Tables \ref{subgroups} and \ref{notsubgroups}).  Of the four GEEC groups with substructure only GEEC Group 320 shows a velocity distribution consistent with being Gaussian.  Five groups, GEEC Groups 4, 38, 138, 238, and 346, are identified as being dynamically complex, but do not contain any substructure according to the DS Test.  

\citet{pinkney96} found that different statistical tools (i.e.\ 1-D, 2-D and 3-D tests) probe the dynamical state of a system at varying epochs.  Using $N$-body simulations, these authors determined that 1-D tests, such as the AD Test are most sensitive to scenarios when substructure passes through the core of the host group (i.e.\ core-crossing).  During this time substructure can become spatially mixed within the host group, and if the substructure is loosely bound then it may be difficult to detect with 3-D tests, such as the DS Test.  Thus, groups with non-Gaussian velocity distributions may contain substructure that is missed by the DS Test.  

\subsection{Comparison with the Velocity Dispersion Profiles}
\indent{}In a study of the VDPs of galaxy clusters, \citet{mff96} found a correlation between the efficiency of merger activity in the cluster core and the shape of the VDP, where actively interacting systems had strongly rising profiles.  We presented a similar correlation for galaxy groups in \citet{hou09}, where we found that groups classified as dynamically complex (i.e.\ non-Gaussian velocity distributions) also had rising VDPs.  We now look at the VDPs of the current GEEC group sample to see if there is a similar relationship between the shape of the profile and the detection of substructure.    

The VDPs are computed following the method outlined in \citet{bergond06}.  Unlike traditional methods of computing binned projected velocity dispersions this technique generates a smoothed VDP.  This is done by using a `moving window' prescription, which takes into account the contribution of every radial velocity measurement at each computed radius.  The values are binned with an exponentially weighted moving window given by
 \begin{equation}
 \centering
 w_{i}(R) = \frac{1}{\sigma_{R}}\exp\left[\frac{(R-R_{i})^{2}}{2\sigma_{R}^{2}}\right]
\label{vdpwindow}
\end{equation}
where $\sigma_{R}$ is the width of the window, which can be constant or a function of radius \emph{R}, and the $R_{i}$'s are the radial positions of the members of the system.  The projected velocity dispersions are then defined as:
 \begin{equation}
 \centering
  \sigma_{p}(R) = \sqrt{\frac{\sum_{i}w_{i}(R)(x_{i} - \bar{x})^{2}}{{\sum_{i}w_{i}(R)}}}
\label{sigmap}
 \end{equation}
where the $x_{i}$'s are the radial velocities and $\bar{x}$ is the mean velocity of the system.

\begin{figure}
\centering
\includegraphics[width = 8.7cm, height = 8.7cm]{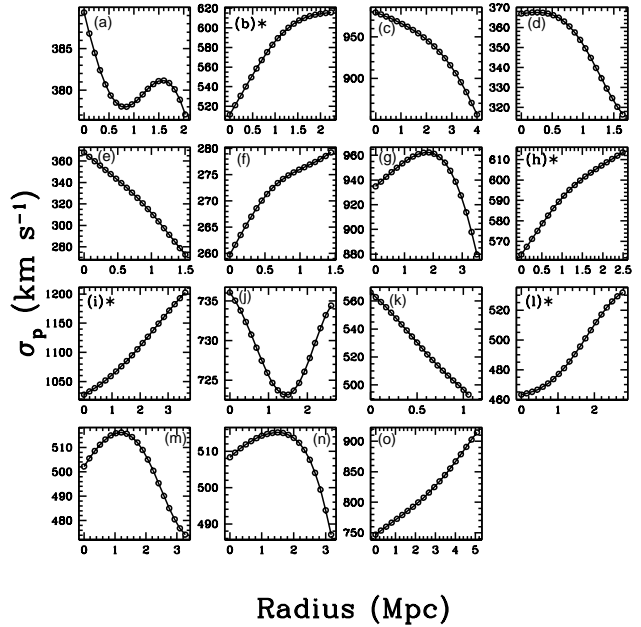}
\caption{Velocity dispersion profiles (VDPs) for the GEEC Groups with $n \geq 20$: panel (a) Group 4, (b) Group 25, (c) Group 38, (d) Group 104, (e) Group 110, (f) Group 117, (g) Group 138, (h) Group 208, (i) Group 226, (j) Group 238, (k) Group 308, (l) Group 320, (m) Group 334, (n) Group 346 and (o) Group 362.    The plots with asterisks in the top-left corner indicate groups that have been identified as having significant substructure.  The intrinsic velocity dispersions, with errors, for each group can be found in Tables \ref{subgroups} and \ref{notsubgroups}.}
\label{vdps}
\end{figure}

We compute VDPs for our GEEC group sample and the profiles are shown in Fig. \ref{vdps} using a widow width, $\sigma_{R}$, equal to one-third the maximum group radius.  It should be noted that the projected velocity dispersions do not include any redshift or instrumental error corrections, so the intrinsic dispersion values (Tables \ref{subgroups} and \ref {notsubgroups}) are generally lower than the projected velocity dispersions.  

Comparing the profiles of the groups with and without substructure, we find that all four GEEC groups identified as having substructure (Groups 25, 208, 226 and 320) also have strongly rising profiles.  In contrast, almost all of the groups with no detected substructure, with the exception of GEEC Group 362, either have flat or generally decreasing VDPs, within the intrinsic velocity dispersion error.  See Tables \ref{subgroups} and \ref{notsubgroups} for a description of the shape of the VDP for each GEEC group.  Thus, we do indeed observe a correlation between detectable substructure and rising VDPs.  As previously mentioned, studies of rich galaxy clusters suggest that a strongly increasing profile may be a signature of merger activity or galaxy interactions \citep{mff96}, but an alternative explanation of rising VDPs is the presence of subclumps with different mean velocities \citep{girardi96, barrena07}.  In our case, it is likely that the increasing profile is being caused by the kinematically distinct substructure, which has a different mean velocity from the host group.     

\section{Correlations between Substructure and Galaxy Properties}
\indent{}Thus far, we have looked at possible correlations between substructure and other indicators of the dynamical state of a system.  We now compare properties of the member galaxies to see if there are any differences between the galaxies in groups with and without substructure.

\subsection{Substructure and Colour}
\indent{}In the following analysis we compare the $^{0.4}(g - r)$ colours, which have been corrected for galactic extinction and k-corrected to a redshift of $z = 0.4$ \citep{balogh09}, and blue fractions,  $f_{b}$, for the groups with and without substructure.

In addition to extinction and k-corrections, we also apply a completeness correction to address the differing spectroscopic coverage.  We apply magnitude weights that depend on whether the group had follow-up spectroscopy.   The weights we apply are similar to those derived in \citet{wilman05}, except we do not include any radial weights.  We compute weights in $r$-band magnitude bins of 0.25, and up to a limit of $r = 22.0$, which is the limit of the unbiased Magellan spectroscopy.  These weights are then applied to all of the member galaxies in our sample.

\begin{figure}
\includegraphics[width = 4.3cm, height = 4.3cm]{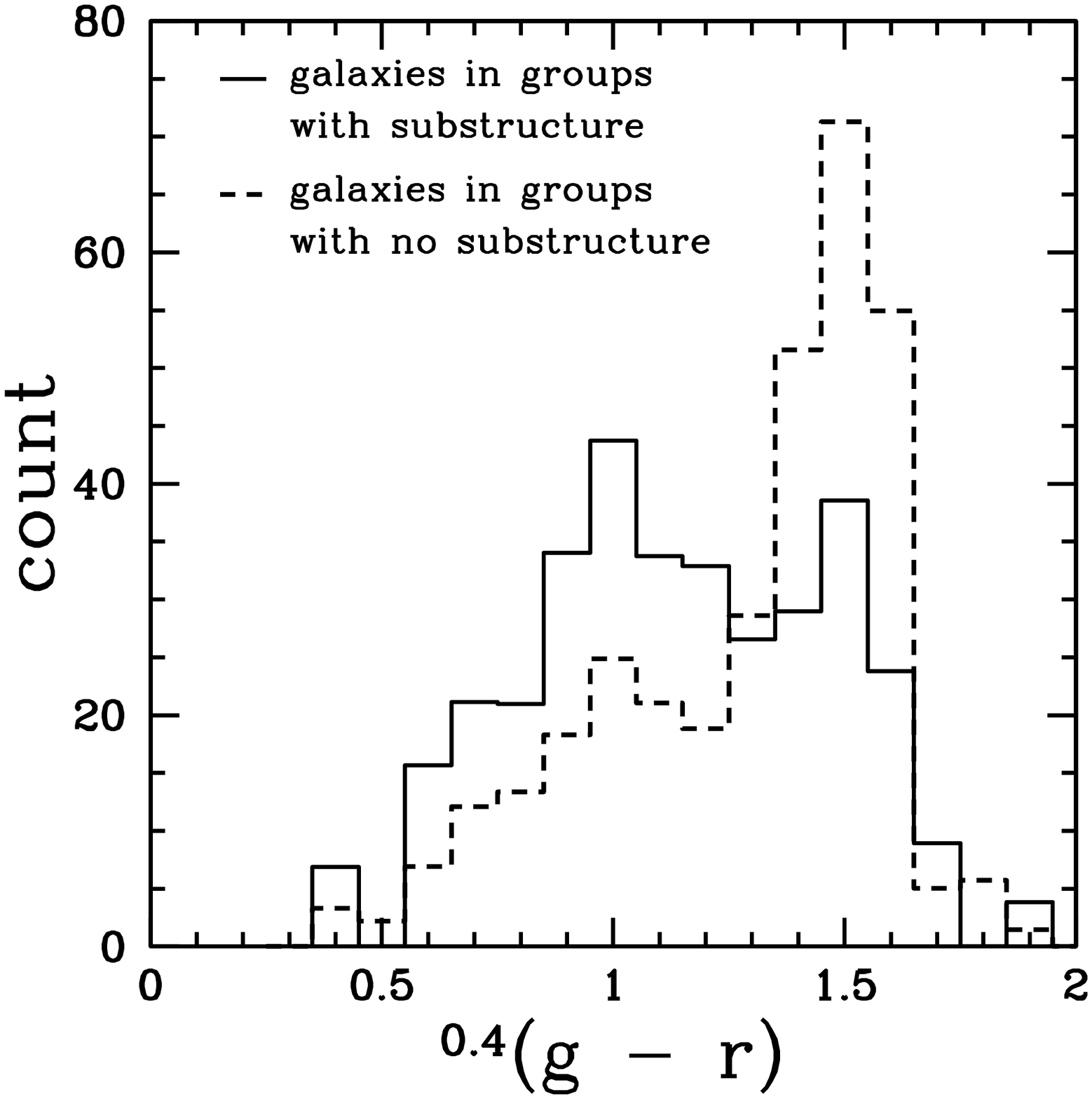}
\includegraphics[width = 4.3cm, height = 4.3cm]{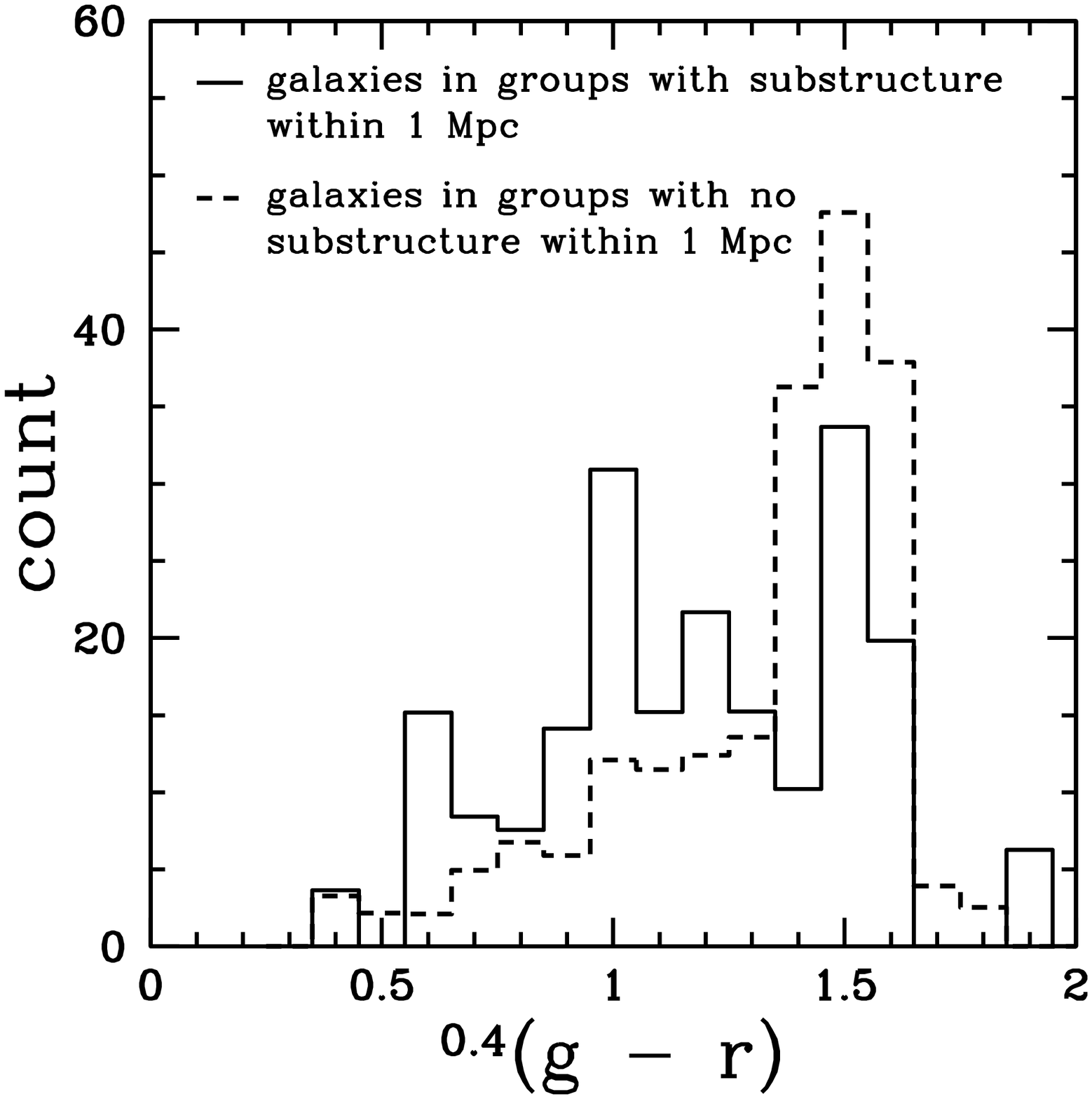}
\caption{Left: $^{0.4}(g - r)$ colour histograms for the GEEC groups with substructure (solid line),  which has be normalized to match the number count of the groups with no substructure and for the GEEC groups with no detectable substructure (dashed line).  Both histograms have r-band magnitude based weights which have been computed to take into account the differing spectroscopic coverage between the original CNOC2 survey and the follow-up Magellan survey.  Right:  Same as figure on the left, expect a 1.0 Mpc radial cut has been applied to all groups in the sample.}
\label{grhist_subnosub}
\end{figure}

In Fig. \ref{grhist_subnosub} we show the completeness weighted $^{0.4}(g - r)$-histograms for the galaxies, with $r < 22.0$, in groups with substructure (solid line) and for those in groups with no detected substructure (dashed line).  From this figure, it is clear that both histograms are bimodal with a distinct red sequence and blue cloud.  Although both colour distributions have the expected bimodal shape, it is obvious that galaxies in the groups with and without substructure come from very different parent colour distributions.  The groups with no identified substructure have a well populated red sequence, while the groups with substructure appear to have a much more dominant blue cloud.  Results from a two-sample Kolmogorov-Smirnov (KS) Test on the unbinned $^{0.4}(g - r)$ distributions show that these two samples are distinct at the 99 per cent c.l.

We also plot colour distributions of the groups with and without substructure with a 1.0 Mpc radial cut applied to all groups in our sample (Fig. \ref{grhist_subnosub}: right).  Although, the blue cloud for the groups with substructure in less populated when a radial cut is applied, it is still more populated than the groups without substructure and the two distributions are still statistically distinct.  This result indicates that the increase in blue galaxies in the groups with substructure is not only coming from galaxies at radii greater than 1.0 Mpc.  Our findings are similar to those of \citet{ribeiro10} who showed that there are many more red galaxies in dynamically evolved group systems, even out to 4 virial radii. 

\begin{figure}
\centering
\includegraphics[width = 4.3cm, height = 4.3cm]{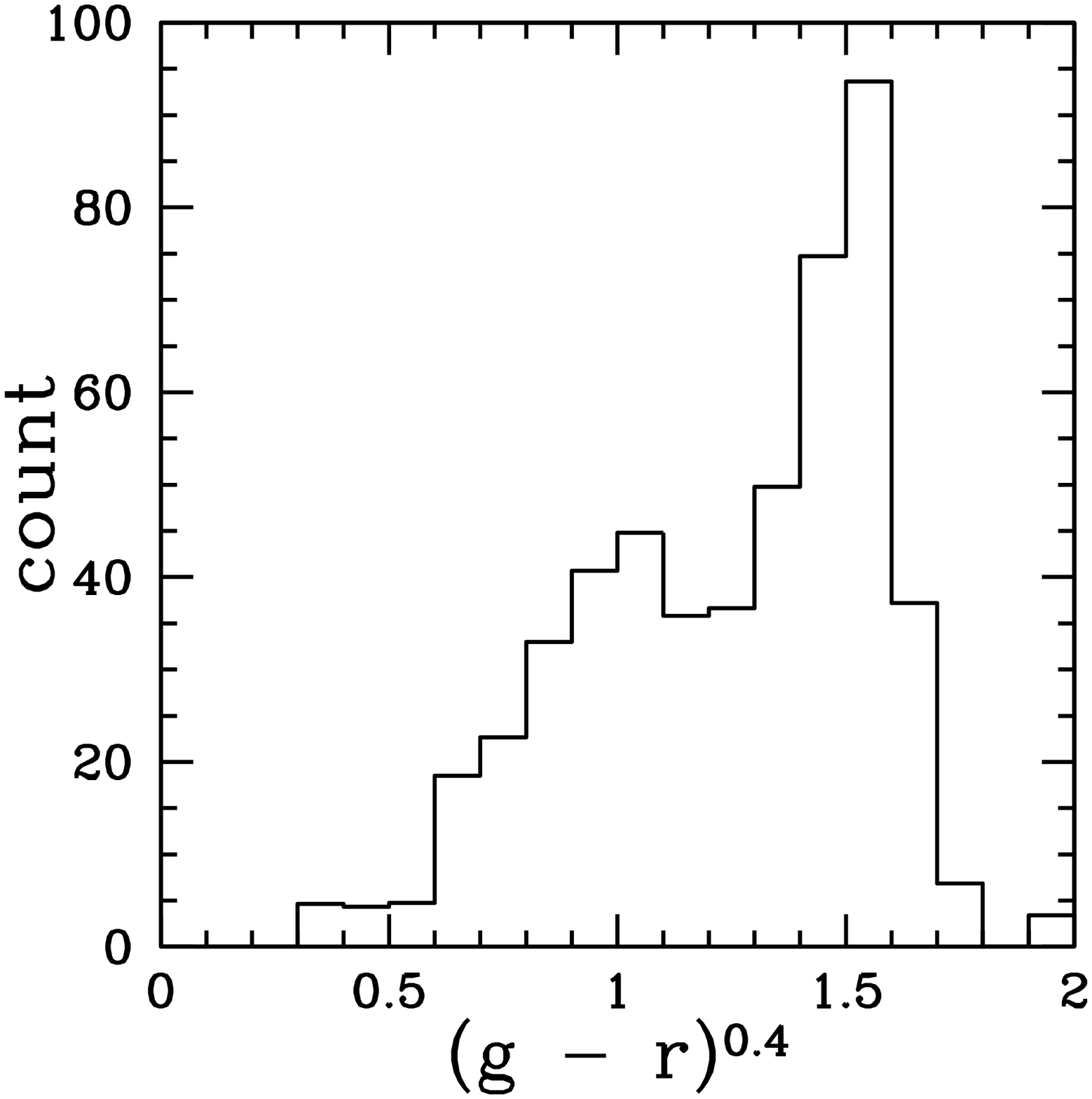}
\includegraphics[width = 4.3cm, height = 4.3cm]{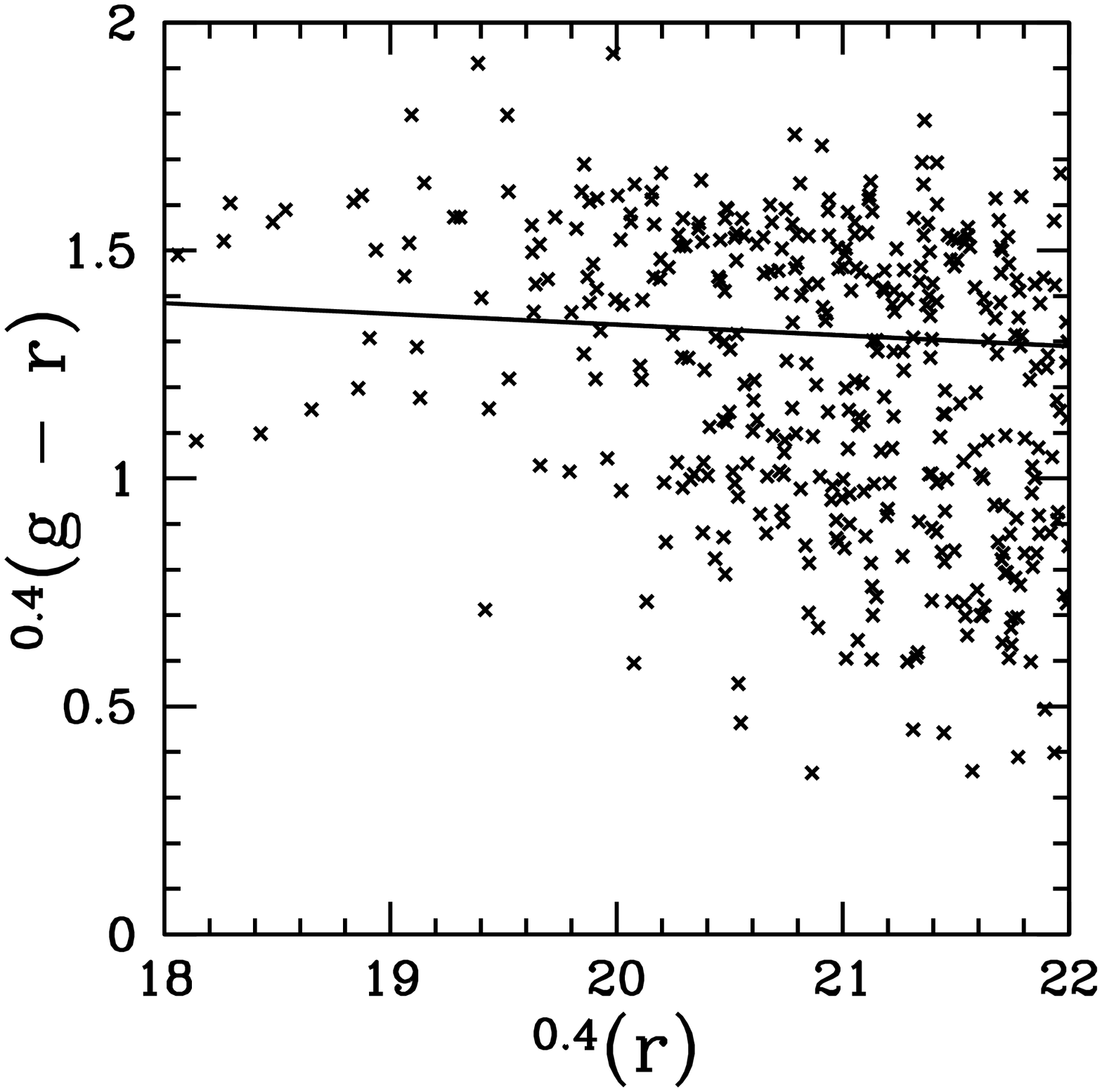}
\caption{Left: Weighted $^{0.4}(g - r)$ histogram of all the member galaxies in our group sample.  Right: $^{0.4}(g - r)$ versus $r$ colour-magnitude diagram of all the galaxies in our sample, with no completeness correction.  The solid line indicates the colour cut used to distinguish between the red sequence and the blue cloud.}
\label{CMD}
\end{figure}

To obtain a more quantitative comparison we compute the fraction of blue galaxies within each sample.  We take a multi-stage approach to determine the appropriate colour cut needed to distinguish between the red sequence and blue cloud.  First, we apply an initial colour cut of $^{0.4}(g - r) = 1.2$, based on the minimum value in the colour distribution of all the member galaxies in our sample, shown in Fig. \ref{CMD}.  We then determine a linear fit to all of the galaxies with $^{0.4}(g - r) > 1.2$ and set the colour cut to be one standard deviation below the fit to the red sequence.  The final colour cut is determined to be
\begin{equation}
 \centering
^{0.4}(g - r) = -0.0236(^{0.4}r) + 1.810,
\label{ccut}
\end{equation}
and is represented by the solid line in the colour-magnitude diagram shown in Fig. \ref{CMD}. 

The blue fraction, $f_{b}$, is then computed as the ratio of galaxies with $^{0.4}(g - r)$ values that fall below Equation \ref{ccut} to the total number of galaxies in the sample.  The error in the blue fraction is computed using confidence intervals (CIs) derived from the beta distribution \citep{cameron10}.  This method has been shown to more accurately determine CIs, especially for small samples, over traditional methods such as Poisson errors, which systematically under-estimates the width of the CIs.  We find that $f_{b} = 41 \pm ^{3}_{4}$ per cent for the groups without substructure and $f_{b} = 69 \pm ^{4}_{6}$ per cent for the groups with substructure.  Thus, the groups with substructure have a significantly higher blue fraction, which is clear in the $^{0.4}(g - r)$-histograms of Fig. \ref{grhist_subnosub}.  We note that although these blue fractions are derived from weighted colour distributions, they are in agreement with the unweighted values,\footnote{$f_{b} = 44 \pm 4$ per cent for the groups without substructure and $f_{b} = 68 \pm ^{5}_{6}$ per cent for the groups with substructure} suggesting that the applied magnitude weights do not affect the observed differences in colour.

In addition, we also compare the blue fractions of the individual galaxies identified as being part of the local substructure (see Section 3.3) to the other members of the group.  We find that for the galaxies in the identified regions of localized substructure,  $f_{b} \simeq 74 \pm ^{8}_{10}$ per cent, and for the galaxies not in the substructure,  $f_{b} \simeq 50 \pm ^{3}_{4}$ per cent.  This suggests that the observed increase in the blue fraction of groups with substructure is being enhanced by the galaxies in the identified regions of local substructure.

\begin{figure}
\centering
\includegraphics[width = 8cm, height = 8cm]{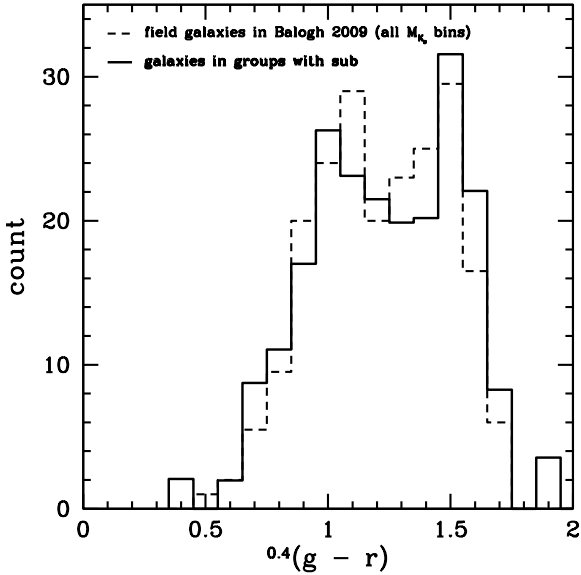}
\caption{$^{0.4}(g - r)$ colour histograms for the GEEC groups with substructure (solid line) and for all of field galaxies in the GEEC groups found in Fig. 6 of \citet{balogh09} (dashed line).  It should be noted that the colour distribution for the field galaxies is a sum of all the counts in each magnitude bin of Fig. 6 of \citet{balogh09}, since our group galaxy sample covers the entire magnitude range.  In addition, the group sample has been normalized to match the number count of the field sample.}
\label{submlb}
\end{figure}

We now compare our groups with substructure sample to a sample of intermediate redshift field galaxies.  In Fig. \ref{submlb} we reproduce the $^{0.4}(g - r)$-histogram for the field galaxies in Fig. 6 of \citet{balogh09} (dashed line), summing up all of the counts in each of the quoted $M_{K_{s}}$ bins in order to get the total colour distribution\footnote{It should be noted that although only a fraction of our sample actually have measured $M_{K_{s}}$ values, those that do span the entire range of magnitudes quoted in Fig. 6 of \citet{balogh09}}.  We also over-plot the colour histogram for the galaxies in groups with substructure, except we now apply an $M_{i}$ magnitude cut in order to match the $M_{K_{s}}$ range used in \citet{balogh09}.   Both the field galaxies in the \citet{balogh09} sample and the galaxies in our groups with substructure lack a prominent red sequence, and have well populated blue clouds.  Despite subtle differences in the two colour histograms, a two-sample KS Test shows that the $^{0.4}(g - r)$ colours of the field galaxies and groups with substructure galaxies very likely come from the same parent distribution ($P$-value = 0.67).

Our blue fractions can be compared to those computed in the zCOSMOS survey, where \citet{iovino10} determined the blue fractions of isolated and group galaxies at various redshifts in their sample.  At a redshift of $z = 0.4$, \citet{iovino10} found $f_{b} \sim 70$ per cent for isolated galaxies and $f_{b} \sim 45$ per cent for group galaxies\footnote{We look at the Sample II of \citet{iovino10}, as this is the data-set that corresponds best to our GEEC sample.}.  From these results it is clear that our observed blue fraction of 69 per cent for the galaxies in groups with substructure is significantly higher than the observed zCOSMOS group sample, but is in agreement with their field sample.  On other hand, the $f_{b}$ values for our GEEC groups with no substructure roughly agree with the zCOSMOS group sample.  

\subsection{Substructure as a Function of Colour, Stellar Mass and Star Formation Rates}
\indent{}In the previous sections we compared the $^{0.4}(g - r)$ colours of the galaxies in groups with substructure to those in groups with no detected substructure.  Here we again compare the colours of the member galaxies but now as a function of stellar mass and specific star formation rate (defined as the ratio of the star formation rate to the stellar mass - SSFR).  Additionally, we compare the SSFR distributions of the galaxies in groups with and without substructure.  It should be noted that since only 3 of the 4 original CNOC2 fields were targeted with GALEX, the sample used in the following analysis contains fewer galaxies than used in the colour analysis of Section 4.3.  In our sample of 15 ($n_{\rm{members}} \geq 20$) GEEC groups,  275 group galaxies have measured SFRs, stellar masses and colours, while 401 group members have well determined colours.

The stellar masses and star formation rates (SFRs) for the GEEC sample were obtained from spectral energy distribution (SED) template-fitting to all of the available photometry.  Detailed discussion of the methodology is presented in \citet{mcgee11}, but we give a brief summary here.  The photometry used in the SED-fitting process typically included photometry in the $K$, $i$, $r$, $g$, $u$, NUV and FUV bands \citep[see][for details of the GEEC photometry]{balogh09}.  The observed photometry was then compared to a large grid of model SEDs constructed using the \citet{bc03} stellar population synthesis code and assuming a Chabrier initial mass function.  Following the methodology of \citet{salim07}, \citet{mcgee11} created a grid of models that uniformly sampled the allowed parameters of formation time, galaxy metallicity, and the two-component dust model of \citet{cf00}.  The star formation history was modelled as an exponentially declining base rate with random bursts of star formation of varying duration and relative strength.  The model magnitudes were obtained by convolving these model SEDs with the observed photometric bandpasses at nine redshifts between 0.25 and 0.60.  $\chi^{2}$-minimization was then performed by summing over all of the models and taking into account the observed uncertainty on each point.  The one sigma uncertainties in stellar mass, when compared to both mock groups and other independent estimates, are on the order of 0.15 dex and the SFRs have been averaged over the last 100 Myr \citep{mcgee11}.  It should be noted that there may be additional systematic uncertainties due to, for example, the Initial Mass Function assumed in the fitting procedure.
 
\begin{figure}
\includegraphics[width = 8.5cm, height = 8.5cm]{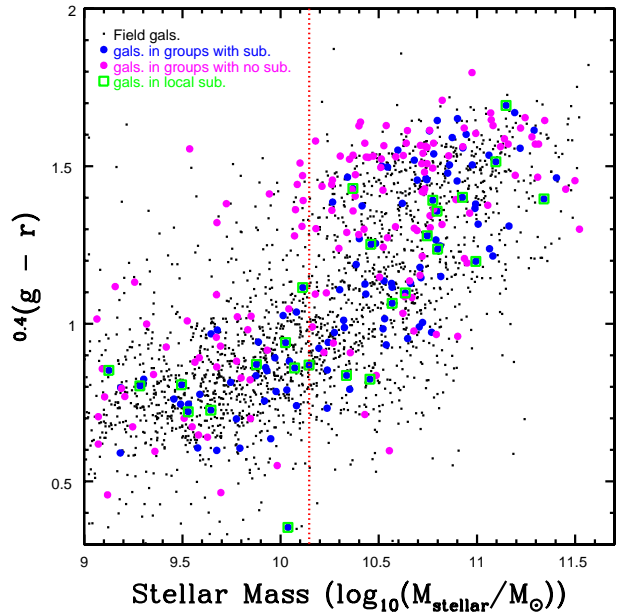}
\caption{$^{0.4}(g - r)$ versus stellar mass for field galaxies (black dots), galaxies in groups with substructure (blue circles), galaxies in groups with no substructure (magenta circles) and galaxies identified as part of localized substructure (green squares).  The red dotted line indicates the stellar mass limit ($\text{M}_{\text{stellar}} = 1.4 \times 10^{10} \text{M}_{\odot}$) at median redshift (z $\sim$ 0.3) of the GEEC sample.}
\label{gr_smass}
\end{figure}

In Fig. \ref{gr_smass} we show $^{0.4}(g - r)$ versus stellar mass for the field galaxies (black dots), galaxies in groups with substructure (closed blue circles), galaxies in groups with no identified substructure (closed magenta circles) and the galaxies identified as being part of local substructure (open green squares).  The dotted red vertical line in Fig. \ref{gr_smass} represents the stellar mass limit of 1.4 $\times 10^{10} \text{M}_{\odot}$ at the median redshift ($z \sim$ 0.3) in the GEEC group sample \citep{mcgee11}.  It should be noted that this stellar mass limit shifts to lower(higher) masses for groups at lower(higher) redshifts.  From Fig. \ref{gr_smass}, it is clear that the galaxies in groups with substructure lie preferentially along the blue cloud, as discussed in Section 4.3, but we also see that these galaxies span a similar mass range as the galaxies in groups with no identified substructure.  A two-sample KS Test of the stellar mass distributions of the galaxies with and without substructure shows that the two distributions likely come from the same parent distribution ($P$-value = 0.55)  The similar stellar mass distribution of the two populations tells us that the DS Test does not simply detect substructure in groups which are probed further down the stellar mass function.

Now we examine the SSFRs for groups with and without substructure.  Following \citet{mcgee11}, we define actively star-forming galaxies to have ${\log_{10}(\text{SFFR}) > -11}$ and the fraction of actively star-forming galaxies to be ${f_{\text{active}} = n_{\text{actively star-forming galaxies}}/n_{\text{total}}}$.  We find that for galaxies in groups with substructure $f_{\text{active}} = 63 \pm 8$ per cent and for groups with no detected substructure $f_{\text{active}} = 49 \pm ^{6}_{5}$ per cent.  These active fractions agree with the blue fractions found in Section 5.1.  

\begin{figure}
\includegraphics[width = 8.5cm, height = 8.5cm]{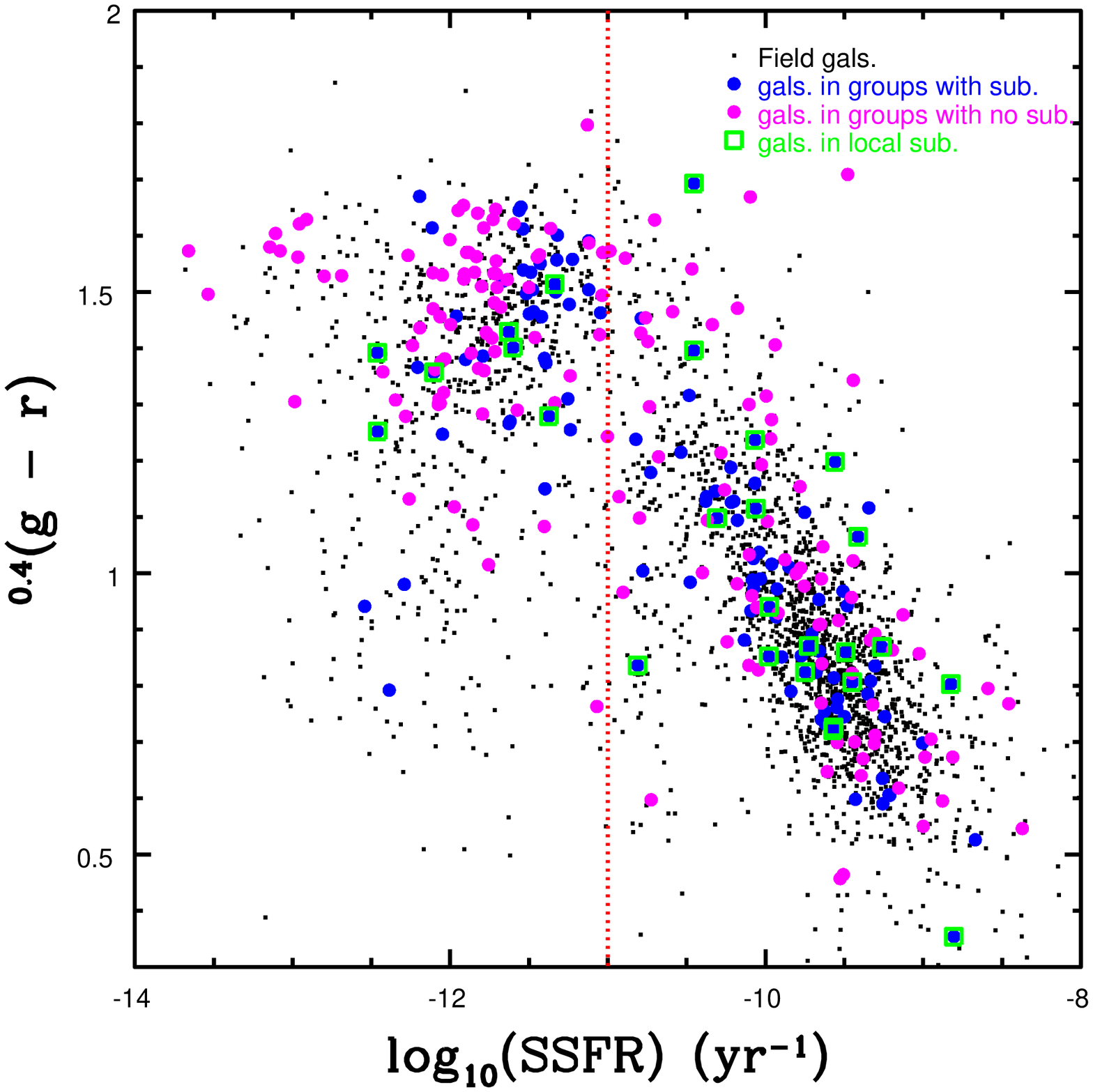}
\newline
\includegraphics[width = 8.5cm, height = 8.5cm]{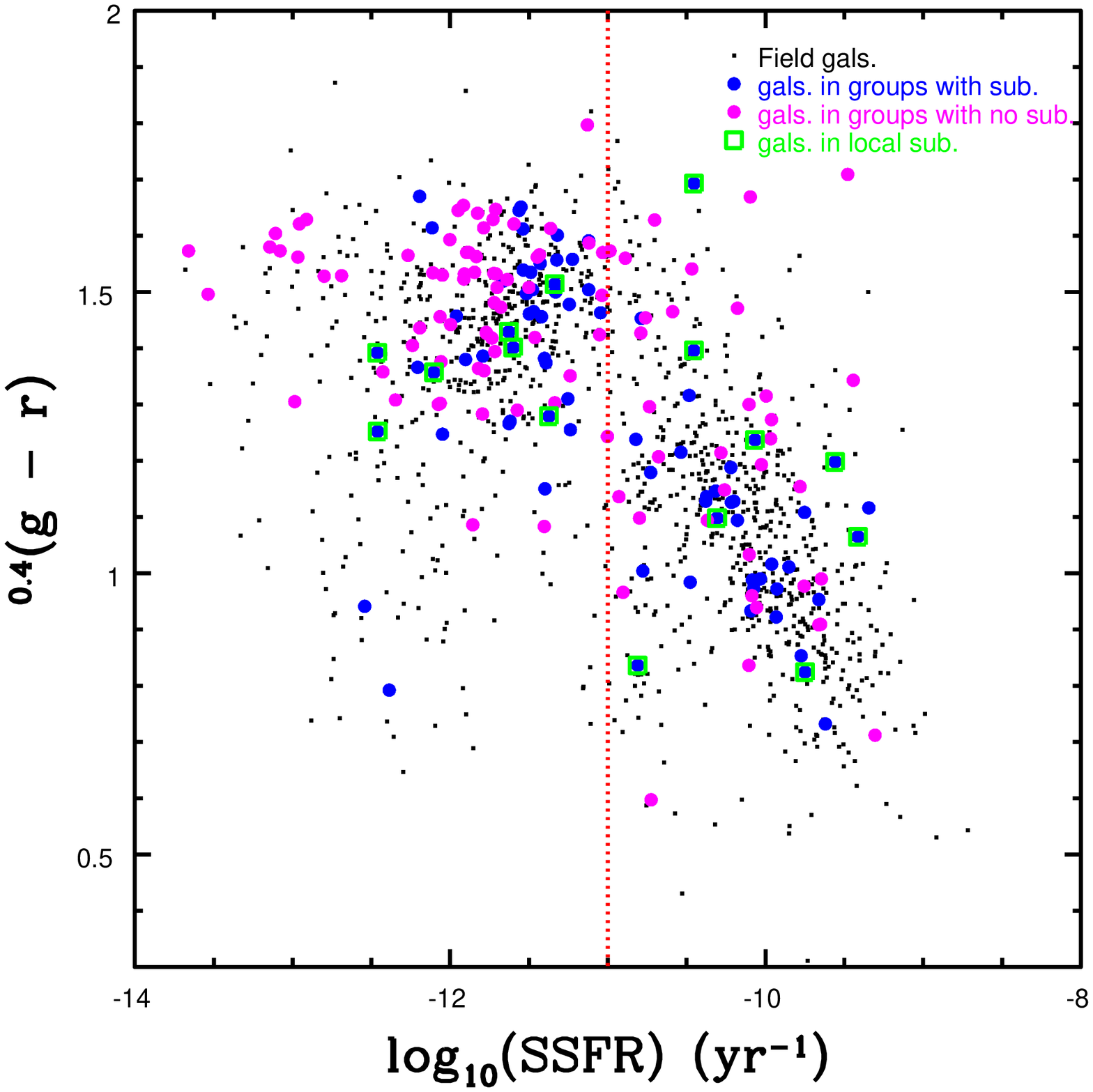}
\caption{Top: $^{0.4}(g - r)$ versus specific star formation rate (SSFR) for field galaxies (black dots), galaxies in groups with substructure (blue circles), galaxies in groups with no substructure (magenta circles) and galaxies identified as part of localized substructure (green squares).  The red dotted line at $\log_{10}(\text{SFFR})$ yr$^{-1}$ indicates the division between active ($\log_{10}(\text{SFFR}) > -11$ yr$^{-1}$) and passive ($\log_{10}(\text{SFFR}) < -11$ yr$^{-1}$) galaxies determined in \citet{mcgee11}. Bottom: Same as figure on the left except with a M$_{\text{stellar}} > 1.4 \times 10^{10} \text{M}_{\odot}$ stellar mass cut applied.}
\label{gr_ssfr}
\end{figure}

Both blue and active fractions are used as independent indicators of quiescent versus actively star-forming galaxies.   However, colour and SSFR probe significantly different timescales, and might be telling us something different about the star formation history of galaxies.  For instance, a dust enshrouded star-forming galaxy would be classified as `red', and therefore quiescent, based on colour but would be classified as actively star-forming based on SSFR.  A better approach is to look at colour and SSFR simultaneously \citep{weinmann06}.  We follow this approach in Fig. \ref{gr_ssfr} where we plot $^{0.4}(g - r)$ versus SSFR for all the galaxies in our sample (top) and with a M$_{\text{stellar}} > 1.4 \times 10^{10} \text{M}_{\odot}$ cut applied to the sample (bottom).  The colour scheme is the same as in Fig. \ref{gr_smass}, except the red dotted line now corresponds to the division between actively star-forming and quiescent/passive galaxies.  From Fig. \ref{gr_ssfr}, we can see that there is a correlation between colour and SSFR with two well populated regions of the plot that correspond to `red and passive' and `blue and active' galaxies, where active refers strictly to actively star-forming galaxies.  In Table \ref{grssfrfracs}, we list the percentage of all galaxies in our sample that populate each region of the colour-SSFR space in Fig. \ref{gr_ssfr} with errors computed using the methodology described in \citet{cameron10}.  Similarly, Table \ref{grssfr_smlim} lists the same information but for galaxies above the stellar mass completeness limit of $1.4 \times 10^{10} \text{M}_{\odot}$.  From these tables we see that the galaxies in groups with substructure have significantly more blue and actively star-forming galaxies than groups with no substructure, though slightly less than the fraction observed in the field.  This result indicates that environmentally driven mechanisms of star-formation quenching are not as efficient in groups with observed substructure.   We note that although the percentages within each region of colour-SSFR space differ between the whole versus stellar mass limited sample, the general trends remain the same.  Since the mass-selected sample only includes galaxies with the highest stellar masses, we expect a decrease in the `blue and active' region due to stellar mass trends (i.e.\ higher mass galaxies are preferentially more red and passive) \citep{iovino10, peng10}.

While the majority of galaxies are either `red and passive' or `blue and active', a non-negligible fraction appear to lie in the other two regions of Fig. \ref{gr_ssfr}.  Even if we take into account the small uncertainties in colour (typically 0.02 mags) and the uncertainties in SSFR \citep[][quote one sigma errors on the order of 0.25 dex]{mcgee11}, a measurable fraction of galaxies still remain in these two regions.  From Table \ref{grssfrfracs}, we see that $\sim$9 per cent of all galaxies in the field and group samples reside in the `blue and passive' region.  This value is higher than the $\sim$1.1 per cent observed by \citet{weinmann06}, though appears similar to the results of \citet{laralopez10}.  Note that both these results are based on SDSS galaxies.  Although we do observe a substantial population of `blue and passive' galaxies in Fig. \ref{gr_ssfr}, we acknowledge that obtaining accurate measures of SSFR for galaxies with low SFRs is notoriously difficult.  As discussed in \citet{mcgee11} the errors in the measured SSFRs increase for galaxies with lower values and some of these galaxies may have underestimated SSFRs.

The final region of the $^{0.4}(g  - r)$-SSFR plot in Fig. \ref{gr_ssfr} corresponds to `red and active' galaxies.  While this population is negligible for the groups with substructure sample and small for the field sample (see Table \ref{grssfrfracs}), it contains $\sim$11$\pm^{4}_{3}$ per cent of the galaxies in the groups with no detected substructure and $\sim$14$\pm^{6}_{4}$ for the stellar mass limited sample.  Although this population may be a result of dusty star-forming galaxies or edge-on discs with strong extinction, it seems unlikely that these galaxies would preferentially be found in groups with no substructure.  An alternative explanation is that this population could be the `transition' galaxies observed by \citet{wolf09}.  These galaxies are still star-forming but at a lower rate ($\sim$4 times) than the field and have more obscured star formation resulting in weak optical signatures (i.e.\ not blue).  A population of `transition' galaxies could explain why there are more `red and active' galaxies in groups with no substructure, as a relaxed system could contain galaxies that are being quenched but have not had their star formation completely cut off.  

\begin{table*}
\centering
\caption{Percentage of all galaxies in our sample within a given region of $^{0.4}(g - r)$ versus SSFR space \label{grssfrfracs}}
\vspace{0.5cm}
\begin{tabular}{cccc}
\hline\hline
$^{0.4}(g - r)$-SSFR region & Galaxies in groups with & Galaxies in groups with & Field galaxies\\
 & substructure &  no substructure & \\
\hline
red and passive$^{a}$ &  $28\pm6$ & $42\pm^{5}_{6}$ & $17\pm1$\\
red and active$^{b}$ & $3\pm^{3}_{2}$ & $11\pm^{4}_{3}$ & $4\pm1$\\
blue and passive & $9\pm^{4}_{3}$ & $9\pm^{4}_{3}$ & $9\pm^{2}_{1}$\\
blue and active & $60\pm^{7}_{6}$ & $38\pm^{6}_{5}$ & $70\pm^{2}_{1}$\\
\hline
\end{tabular}
\newline
$^{a}$Passive denotes quiescent galaxies with $\log_{10}(\text{SFFR}) < -11$ yr$^{-1}$\\
$^{b}$Active refers to actively star-forming galaxies with $\log_{10}(\text{SFFR}) > -11$ yr$^{-1}$

\caption{Same as Table \ref{grssfrfracs} except for galaxies above the stellar mass completeness limit of M$_{\text{stellar}} > 1.4 \times 10^{10} \text{M}_{\odot}$ \label{grssfr_smlim}}
\vspace{0.5cm}
\begin{tabular}{cccc}
\hline\hline
$^{0.4}(g - r)$-SSFR region & Galaxies in groups with & Galaxies in groups with & Field galaxies\\
 & substructure &  no substructure & \\
\hline
red and passive &  $42\pm8$ & $57\pm7$ & $33\pm2$\\
red and active & $4\pm^{5}_{2}$ & $14\pm^{6}_{4}$ & $8\pm1$\\
blue and passive & $13\pm^{7}_{4}$ & $8\pm^{5}_{3}$ & $11\pm1$\\
blue and active & $41\pm^{8}_{7}$ & $21\pm^{6}_{5}$ & $48\pm2$\\
\hline
\end{tabular}
\end{table*}

\subsection{Implications of the Observed Properties of Groups with Substructure}
\indent{}The field-like colour distribution of the groups with substructure and the fact that the substructure galaxies are found in the group outskirts may have implications for the nature of environmental effects in galaxy evolution.  It is well known that the properties of galaxies depend, at least in some part, on their local environment \citep{pg84, dressler97}.  Galaxies that reside in dense environments, such as groups or clusters, generally experience some form of star-formation attenuation due to processes such as ram pressure stripping \citep{gg72, abadi99, quillis00}, strangulation \citep{ltc80, balogh00, km08} or mergers and interactions \citep{tt72, brough06}.  However, the precise details of the galaxy transformation process (i.e.\ exactly when and where quenching occurs, which mechanisms dominate in the different environment, etc.) are still unclear.  

Our substructure analysis in the GEEC groups suggests that the identified local substructure in our sample, which in some cases appears to be infalling, do not feel any strong environmental effects from the host group.  The observed colours and SSFR's of the galaxies in groups with substructure are significantly more blue, active and remarkably field-like when compared to galaxies in groups with no substructure.  This suggests that an enhanced fraction of red galaxies - either/both via the suppression of star formation or/and dust obscuration - only happens in relaxed groups with no detected substructure.  Thus, any environmental effects felt by infalling substructure galaxies do not likely occur until well inside the group potential.  

Recent studies of star-formation as a function of group- or cluster-centric radius have produced conflicting results with regards to the radius at which environmental effects become observable.  Similar to our results, \citet{wetzel11} conclude that galaxies do not show suppressed star formation outside the virial radius.  In contrast, \citet{vdl10} state that suppressed star formation could be detected in SDSS clusters out to $\sim$ several virial radii.  Such differences are likely sensitive to group or cluster finding algorithms as well as membership assignment.

In a study of the effects of environment on the colours of galaxies in the SDSS survey, \citet{wilman10} take a different approach to classifying environment.  Rather than using a catalog derived from a group-finding algorithm, these authors parametrize the environment using non-overlapping annular measurements of density on independent scales, allowing for comparison of environmental effects at various radii.  Based on their analysis, \citet{wilman10} concluded that the fraction of red galaxies correlated with local density only up to scales of $\sim$1 Mpc, which is similar to our results, as well as those of \citet{wetzel11}.  Though numerous and independent analyses of environmental effects on galaxy evolution seem to indicate that star-formation truncation does not occur until galaxies are well-inside the group/cluster potential, there are studies that suggest the contrary.  Clearly more work, both observational and theoretical, is needed to better understand when and where environmental effects become observable.   

\section{Conclusions}
\indent{}We have studied the Dressler-Shectman Test for substructure to determine the sensitivity and reliability of this test for group-sized systems.  Using mock groups with and without substructure, generated using Monte Carlo methods, we find that the DS Test can reliably be applied to groups with more than 20 members, \emph{if} the probabilities, or $P$-value, method is used with a high confidence level of 99 or 95 per cent.  We also find that for groups with $10 \leq n_{\rm{members}} < 20$, the DS Test cannot detect all of the substructure within a system, but it can be used to determine a reliable \emph{lower limit} on the amount of substructure.  

Of the 15 rich GEEC groups, with a velocity dispersion range of $\sim$260-950 km s$^{-1}$ and $ 20 < n_{\rm{members}} < 90$, we find that 4 groups are identified as having significant substructure.  Further analysis indicates that 2 of these systems, GEEC Groups 208 and 320, likely have gravitationally bound local substructure that lies on the group outskirts and could be accreting onto the system.  

We then looked at various dynamical and galaxy properties to search for correlations with the presence of substructure.  The main results of this analysis are;
\begin{enumerate}
\item The majority of groups with detected substructure also have non-Gaussian velocity distributions;
\item The shape of a group's velocity dispersion profile (VDP) correlates with the detection of substructure, where GEEC groups with substructure have rising profiles;
\item The $^{0.4}(g -r)$ colour distributions of the groups with and without substructure are found to be significantly different, and the colour distribution for the galaxies in groups with substructure is similar to the field distribution;
\item Groups with substructure have a significantly higher fraction of blue galaxies, as do the galaxies within identified regions of localized substructure;
\item Groups with substructure have a larger fraction of actively star-forming galaxies ($\log_{10}(\text{SFFR}) > -11$ yr$^{-1}$), when compared to groups with no identified substructure;
\item There is a measurable fraction of galaxies that populate the `red and active' region of $^{0.4}(g -r)$-SSFR space and we find that this fraction is significantly higher in groups with no substructure for both the whole and stellar mass limited samples.
\end{enumerate}

In conclusion, we find that a considerable fraction of intermediate redshift galaxy groups contain significant substructure, which suggests that like massive clusters, groups grow hierarchically through the accretion of smaller structures.  The field-like colour distribution and measured SSFRs of the galaxies in groups with substructure, combined with the location of the substructure, suggests that these galaxies are not experiencing any form of environment-related star-formation quenching.  To fully understand the results presented here within the context of galaxy evolution will require the use of sophisticated modelling.  To this end we plan to duplicate this analysis on a sample of semi-analytic groups obtained from GALFORM simulations \citep{bower06} and compare these with our observational results.  With this we hope to be able to better understand the nature of substructure identified by the DS Test.

\section{Acknowledgments}
We would like to thank the CNOC2 team for the use of unpublished redshifts.  A.H, L.C.P, and W.E.H would like to thank the National Science and Engineering Research Council of Canada (NSERC) for funding.

\bibliography{ahou_group_sub_mnras}

\clearpage
\appendix

\section{False Negative Rates in More Detail}
\indent{}In Section 2.2.3, we presented the main results of the effects of changing various input parameters in our mock groups.  Here we present tables detailing the specific false negative rates obtained and also discuss each free parameter in detail.  As discussed in Section 2.2.3, we determine the sensitivity of the DS Test to each of the free parameters in our mock groups ($\sigma_{\text{position}}$, $\epsilon_{\text{position}}$, $\sigma_{\text{redshift}}$, $\epsilon_{\text{redshift}}$, $n_{\text{sub}}$ and $\sigma_{\text{host}}$) by beginning with a `base' mock group (Table \ref{basefneg}), which has a false negative rate of zero per cent.  We then change only one parameter at a time to ensure that any change in the false negative rate can be directly attributed to the altered free parameter.   

\subsection{Angular Size of the Substructure ($\sigma_{\text{position}}$)}
\indent{}In Table \ref{FNsigpos} we present the false negative rates (listed as a percentage) for mock groups with $n_{member}$ values of 10, 15, 20 and 50 as a function of projected angular size on the sky ($\sigma_{\text{position}}$).  The first line in the table indicate the results for our `base' groups, which have $P$-values of either 0 or 1 per cent.  We then increase the value of $\sigma_{\text{position}}$ and from Table \ref{FNsigpos}, we can see that the DS Test reliably identifies substructure with a projected dispersion of \emph{up to} 0.1 Mpc for groups with 10 members, and as large as 0.1 Mpc for groups with more then 15 members.   If we increase $\sigma_{\text{position}}$ to 0.2 Mpc, we find that for groups with roughly 20 members or less, the false negative rates increases dramatically, but still remain very low (1 per cent) for richer groups with 50 members.  The general conclusion from Table \ref{FNsigpos} is that the DS Test is not very sensitive to the size of the substructure and that even for small groups it can identify real substructure that is relatively large ($\sim$ 0.1 Mpc).
\begin{table*}
\caption{False Negative Rates: Dependency on the Angular Size of the Input Substructure ($\sigma_{\text{position}}$) \label{FNsigpos}}
\begin{tabular}{ccccc}
\hline\hline
$\sigma_{\text{position}}$ & $n_{\rm{members}} = 10$  & $n_{\rm{members}} = 15$  & $n_{\rm{members}} = 20$ & $n_{\rm{members}} = 50$\\
Mpc	& & & & \\
\hline
0.01$^{a}$ & 1$^{b}$ & 0 & 0 & 0\\
0.05 & 3 & 1 & 0 & 0\\
0.09 & 4 & 1 & 3 & 0\\
0.1  & 11 & 3 & 2 & 0\\
0.2  & 48 & 26 & 20 & 1\\
0.5  & 83 & 74 & 73 & 27\\
\end{tabular}
\linebreak
$^{a}$These mock groups have the following input parameters; $\epsilon_{\text{position}}$ = 0.5 Mpc, $\sigma_{\text{redshift}} = 100$ km s$^{-1}$, $\epsilon_{\text{redshift}} = 1300$ km s$^{-1}$ and $\sigma_{\text{host}}$ values listed in Table \ref{basefneg}.  Only the $\sigma_{\text{position}}$ parameter is varied for these trials.\\
$^{b}$Values quoted are the false negative rates, given as a percentage, obtained for 100 trials with each set of inputs, using 100 000 mc shuffles.
\end{table*}

\subsection{Location of Substructure in Position Space ($\epsilon_{\text{position}}$)}
\indent{}In Table \ref{FNsubpos} we list the false negative rates for mock groups with $n_{member}$ values of 10, 15, 20 and 50 as a function of the projected radial distance of the substructure with respect to the group centroid ($\epsilon_{\text{position}}$).  It is clear from this Table that the DS Test is quite insensitive to $\epsilon_{\text{position}}$.  In other words, the test can reliably identify substructure that is `close' to the projected group centre, and easily detects structure that lies on the group outskirts.
\begin{table*}
\caption{False Negative Rates: Dependency on the Location of the Input Substructure in Position Space ($\epsilon_{\text{position}}$) \label{FNsubpos}}
\begin{tabular}{ccccc}
\hline\hline
$\epsilon_{\text{position}}$ & $n_{\rm{members}} = 10$ & $n_{\rm{members}} = 15$ & $n_{\rm{members}} = 20$ & $n_{\rm{members}} = 50$\\
Mpc	& & & & \\
\hline
0.001$^{a}$ & 7$^{b}$ & 2 & 0 & 0\\
0.01 & 11 & 2 & 3 & 0\\
0.1  & 6 & 0 & 0 & 0\\
0.2  & 0 & 1 & 0 & 0\\
0.3  & 2 & 0 & 0 & 0\\
0.4  & 1 & 0 & 0 & 0\\
0.5  & 3 & 0 & 0 & 0\\
1.0  & 0 & 1 & 0 & 0\\
\hline
\end{tabular}
\linebreak
$^{a}$These mock groups have the following input parameters; $\sigma_{\text{position}}$ = 0.01 Mpc, $\sigma_{\text{redshift}} = 100$ km s$^{-1}$, $\epsilon_{\text{redshift}} = 1300$ km s$^{-1}$ and $\sigma_{\text{host}}$ values listed in Table \ref{basefneg}.  Only the $\epsilon_{\text{position}}$ parameter is varied for these trials.\\
$^{b}$Values quoted are the false negative rates, given as a percentage, obtained for 100 trials with each set of inputs, using 100 000 mc shuffles.\\
\end{table*}

\subsection{Velocity Dispersion of the Input Substructure ($\sigma_{\text{redshift}}$)}
\indent{}In Table \ref{FNsigvel} we list the false negative rates for mock groups with $n_{member}$ values of 10, 15, 20 and 50 as a function of the velocity dispersion of the input substructure ($\sigma_{\text{redshift}}$).  From Table \ref{FNsigvel} it is evident that the DS Test is also insensitive to this parameter and can reliably detect substructure with a wide range of velocity dispersions for all values of $n_{\rm{members}}$.  Only for groups with a fewer than 20 members and a very large dispersion value of 450 km s$^{-1}$ do the false negative rates go above 5 per cent.  
\begin{table*}
\caption{False Negative Rates: Dependency on the Velocity Dispersion of the Input Substructure ($\sigma_{\text{redshift}}$) \label{FNsigvel}}
\begin{tabular}{ccccc}
\hline\hline
$\sigma_{\text{redshift}}$ & $n_{\rm{members}} = 10$ & $n_{\rm{members}} = 15$  & $n_{\rm{members}} = 20$ & $n_{\rm{members}} = 50$\\
km s$^{-1}$	& & & & \\
\hline
50$^{a}$ & 0$^{b}$ & 0 & 0 & 0\\
100 & 0 & 0 & 1 & 0\\
150 & 1 & 1 & 0 & 0\\
200 & 0 & 0 & 0 & 0\\
250 & 4 & 2 & 0 & 0\\
300 & 3 & 4 & 0 & 0\\
350 & 3 & 2 & 0 & 0\\
400 & 3 & 4 & 1 & 0\\
450 & 13 & 13 & 3 & 0\\
\hline
\end{tabular}
\linebreak
$^{a}$These mock groups have the following input parameters; $\sigma_{\text{position}}$ = 0.01 Mpc, $\epsilon_{\text{position}} = 0.5$ Mpc, $\epsilon_{\text{redshift}} = 1300$ km s$^{-1}$ and $\sigma_{\text{host}}$ values listed in Table \ref{basefneg}.  Only the $\sigma_{\text{redshift}}$ parameter is varied for these trials.\\
$^{b}$Values quoted are the false negative rates, given as a percentage, obtained for 100 trials with each set of inputs, using 100 000 mc shuffles.
\end{table*}

\subsection{Location of Substructure Along the Line-of-Sight ($\epsilon_{\text{redshift}}$)}
\indent{}In Table \ref{FNsigvel} we list the false negative rates for mock groups with $n_{member}$ values of 10, 15, 20 and 50 as a function of the location of the substructure along the line-of-sight ($\epsilon_{\text{redshift}}$).  This parameter is taken to be a displacement (in km s$^{-1}$) of the peak in the velocity distribution of the substructure with respect to the peak of the host's distribution.  Unlike the previously discussed parameters, the DS Test appears to be extremely sensitive to the location of the substructure along the line-of-sight, or more specifically the separation between the main groups velocity distribution and the substructure velocity distributionty.

The first set of entries in Table \ref{FNsigvel} are the false negative rates from mock groups with $\sigma_{\text{host}}$ values set to the average dispersion of observed GEEC groups with similar group membership (Table \ref{basefneg}).  It is clear that if the peaks of the host and substructure velocity distributions are close ($<$ 300 km s$^{-1}$), then the DS Test cannot always identify real substructure.  For groups with fewer than $\sim$ 20 members, the peaks must be at least 900 km s$^{-1}$ apart in order for the false negative rates to fall below 5 per cent.  

We also find that not only is the DS Test sensitive to the $\epsilon_{\text{redshift}}$ parameter, but that the level of sensitivity is dependent on the number of members in the host group.  This is best seen by looking at the false negative rates listed in the second set of values listed in Table \ref{FNsubvel}, where we use a constant value of $\sigma_{\text{host}} =$ 500 km s$^{-1}$ for all values of $n_{\rm{members}}$.  From this section of the table we see that when the velocity distribution of the substructure is located at $1\sigma_{\text{host}}$, the false negative rates are 43, 49, 34 and 4 per cent for mock groups with 10, 15, 20 and 50 members.  At $2\sigma_{\text{host}}$ the rates are 10 per cent for groups with 10 and 15 members and 0 per cent for groups with 20 and 50 members.  This results indicates that for groups with fewer member galaxies, the velocity distributions of the host and substructure must be very distinct in order for the DS Test to detect substructure.  Alternatively, groups with more than 20 members can have substructure galaxies with a velocity distribution embedded within the host distribution and still be identified by the test.  
\begin{table*}
\caption{False Negative Rates: Dependency on the Location of the Input Substructure Along the Line-of-Sight (i.e.\ redshift space) ($\epsilon_{\text{redshift}}$) \label{FNsubvel}.}
\begin{tabular}{ccccc}
\hline\hline
$\epsilon_{\text{redshift}}$ & $n_{\rm{members}} = 10$ & $n_{\rm{members}} = 15$ & $n_{\rm{members}} = 20$  & $n_{\rm{members}} = 50$\\
km s$^{-1}$	& & & & \\
\hline
100$^{a}$ & 91$^{b}$ & 94 & 80 & 43\\
200 & 85 & 81 & 74 & 39\\
300 & 62 & 71 & 72 & 21\\
400 & 53 & 49 & 54 & 7\\
500 & 32 & 42 & 39 & 0\\
600 & 25 & 25 & 18 & 0\\
700 & 10 & 19 & 7 & 0\\
800 & 7 & 9 & 9 & 0\\
900 & 3 & 5 & 4 & 0\\
1000 & 3 & 2 & 4 & 0\\
1100 & 2 & 0 & 0 & 0\\
1200 & 1 & 1 & 0 & 0\\
1300 & 1 & 0 & 0 & 0\\
\hline
100$^{c}$ & 84 & 94 & 87 & 51\\
200 & 83 & 87 & 67 & 43\\
300 & 80 & 81 & 69 & 21\\
400 & 63 & 66 & 48 & 4\\
500 & 43 & 49 & 34 & 4\\
600 & 37 & 43 & 21 & 0\\
700 & 35 & 26 & 13 & 0\\
800 & 21 & 22 & 3 & 0\\
900 & 17 & 8 & 1 & 0\\
1000 & 10 & 10 & 0 & 0\\
1100 & 5 & 15 & 0 & 0\\
1200 & 5 & 2 & 0 & 0\\
1300 & 2 & 5 & 0 & 0\\
\hline
\end{tabular}
\linebreak
$^{a}$These mock groups have the following input parameters; $\sigma_{\text{position}}$ = 0.01 Mpc, $\epsilon_{\text{position}} = 0.5$ Mpc, $\epsilon_{\text{redshift}} = 1300$ km s$^{-1}$ and $\sigma_{\text{host}}$ values listed in Table \ref{basefneg}.  Only the $\sigma_{\text{redshift}}$ parameter is varied for these trials.\\
$^{b}$Values quoted are the false negative rates, given as a percentage, obtained for 100 trials with each set of inputs, using 100 000 mc shuffles.\\
$^{c}$These mock groups have the following input parameters; $\sigma_{\text{position}}$ = 0.01 Mpc, $\epsilon_{\text{position}} = 0.5$ Mpc, $\epsilon_{\text{redshift}} = 1300$ km s$^{-1}$ and a \emph{constant} $\sigma_{\text{host}}$ value of 500 km s$^{-1}$ for all values of $n_{\rm{members}}$.  Again, only the $\sigma_{\text{redshift}}$ parameter is varied for these trials.
\end{table*}

\subsection{Number of Members in the Input Substructure ($n_{\text{sub}}$)}
\indent{}In Table \ref{FNsigvel} we list the false negative rates for mock groups with $n_{member}$ values of 10, 15, 20 and 50 as a function of the number of members in the input substructure ($n_{\text{sub}}$) (see Section 2.2.3 for discussion).
  
\begin{table*}
\caption{False Negative Rates: Dependency on the Number of Members in the Input Substructure ($n_{\text{sub}}$) \label{FNnsub}}
\begin{tabular}{ccccc}
\hline\hline
$n_{\text{sub}}$ & $n_{\rm{members}} = 10$  & $n_{\rm{members}} = 15$  & $n_{\rm{members}} = 20$ & $n_{\rm{members}} = 50$ \\
\hline
3$^{a}$ & 43$^{b}$ & 39 & 54 & 51\\
4 & 1 & 2 & 9 & 16\\
5 &  & 0 & 0 & 3\\
6 &  &  &  & 0\\
7 &  &  &  & 0\\
8 &  &  &  & 0\\
9 &  &  &  & 0\\
\hline
\end{tabular}
\linebreak
$^{a}$These mock groups have the following input parameters; $\sigma_{\text{position}}$ = 0.01 Mpc, $\epsilon_{\text{position}} = 0.5$ Mpc, $\sigma_{\text{redshift}}$ = 100 km s$^{-1}$, $\epsilon_{\text{redshift}} = 1300$ km s$^{-1}$ and $\sigma_{\text{host}}$ values given in Table \ref{basefneg}.  For these simulations we only change the number of members in the substructure ($n_{\text{sub}}$).\\
$^{b}$Values quoted are the false negative rates, given as a percentage, obtained for 100 trials with each set of inputs, using 100 000 mc shuffles.  A null entry indicates that no trials were run with the associated $n_{\text{sub}}$ value.
\end{table*}

\subsection{Velocity Dispersion of the Host Group ($\sigma_{\text{host}}$)}
\indent{}In addition to the above tests, we also check to see if changing the velocity dispersion, and therefore mass, of the host group ($\sigma_{\text{host}}$) affects the observed false negative rates.  We present the results in Table \ref{FNsighost}, where we see that for groups with more that 20 members the dispersion of the host group does not significantly increase the rate of false negatives.  However, for systems with fewer members and larger values of $\sigma_{\text{host}}$ ($>$ 700 km s$^{-1}$), the DS Test is more likely to miss true substructure.  Fortunately, observed groups with 10 or so members do not generally have such high dispersion values.

\begin{table*}
\caption{False Negative Rates: Dependency on the Velocity Dispersion of the Host Group ($\sigma_{\text{host}}$) \label{FNsighost}.}
\begin{tabular}{ccccc}
\hline\hline
$\sigma_{\text{host}}$ & $n_{\rm{members}} = 10$ & $n_{\rm{members}} = 15$ & $n_{\rm{members}} = 20$ & $n_{\rm{members}} = 50$\\
km s$^{-1}$	& & & & \\
\hline
100$^{a}$ & 0$^{b}$ & 0 & 0 & 0\\
200 & 1 & 0 & 0 & 0\\
300 & 0 & 0 & 0 &  0\\
400 & 1 & 0 & 0 & 0\\
500 & 2 & 2 & 0 & 0\\
600 & 10 & 3 & 1 & 0\\
700 & 9 & 10 & 3 & 0\\
800 & 21 & 20 & 3 & 0\\
900 & 22 & 28 & 3 & 0\\
\hline
\end{tabular}
\linebreak
$^{a}$These mock groups have the following input parameters; $\sigma_{\text{position}}$ = 0.01 Mpc, $\epsilon_{\text{position}} = 0.5$ Mpc, $\sigma_{\text{redshift}}$ = 100 km s$^{-1}$ and $\epsilon_{\text{redshift}} = 1300$ km s$^{-1}$.  For these simulations we keep the values of the input substructure constant, but change the velocity dispersion of the host group ($\sigma_{\text{host}}$).\\
$^{b}$Values quoted are the false negative rates, given as a percentage, obtained for 100 trials with each set of inputs, using 100 000 mc shuffles.
\end{table*}

\end{document}